\documentclass[journal,twoside]{IEEEtran}

\usepackage{mathrsfs}
\usepackage[noadjust]{cite}
\usepackage{graphicx,color,overpic,psfrag}
\usepackage{amsmath, amssymb}
\usepackage{latexsym}
\usepackage{bm}
\usepackage{amssymb}
\usepackage{cases}
\usepackage{array}
\usepackage{fancyhdr}
\usepackage{setspace}
\usepackage{subfigure}

\usepackage{url}
\usepackage{algpseudocode}
\usepackage{algorithm}
\usepackage{blkarray}
\usepackage{booktabs}

\usepackage{multirow}
\usepackage{dsfont}
\usepackage{tabularx}
\usepackage[table]{xcolor}

\usepackage{amsfonts}
\usepackage{letltxmacro}

\graphicspath{{figure/}}

\makeatletter

\newcommand{\Rmnum}[1]{\expandafter\@slowromancap\romannumeral #1@}
\makeatother

\newtheorem{prop}{Proposition}
\newtheorem{proof}{Proof}

\newcommand{\figref}[1]{Fig. \ref{#1}}

\newcommand{\alref}[1]{\textbf{Algorithm \ref{#1}}}
\newcommand{\appref}[1]{Appendix \ref{#1}}
\newcommand{\secref}[1]{Section \ref{#1}}

\newcommand{\propref}[1]{Proposition \ref{#1}}

\newcommand{\Exp}{{\mathsf{E}}}
\newcommand{\expect}[1]{\Exp\left\{#1\right\}}

\newcommand{\tr}[1]{\mathsf{tr}\left\{#1\right\}}
\newcommand{\diag}[1]{\mathsf{diag}\left\{#1\right\}}

\newcommand{\cK}{\mathcal{K}}
\newcommand{\cL}{\mathcal{L}}

\newcommand{\cO}{\mathcal{O}}
\newcommand{\cP}{\mathcal{P}}

\newcommand{\cT}{\mathcal{T}}
\newcommand{\cU}{\mathcal{U}}

\newcommand{\bn}{\mathbf{n}}

\newcommand{\bx}{\mathbf{x}}

\newcommand{\bA}{\mathbf{A}}

\newcommand{\bD}{\mathbf{D}}

\newcommand{\bF}{\mathbf{F}}
\newcommand{\bG}{\mathbf{G}}
\newcommand{\bH}{\mathbf{H}}
\newcommand{\bI}{\mathbf{I}}

\newcommand{\bK}{\mathbf{K}}

\newcommand{\bQ}{\mathbf{Q}}
\newcommand{\bR}{\mathbf{R}}

\newcommand{\bT}{\mathbf{T}}
\newcommand{\bU}{\mathbf{U}}
\newcommand{\bV}{\mathbf{V}}

\newcommand{\bX}{\mathbf{X}}
\newcommand{\bY}{\mathbf{Y}}

\newcommand{\bzero}{\mathbf{0}}

\newcommand{\bPsi}{{\boldsymbol\Psi}}

\newcommand{\bLambda}{{\boldsymbol\Lambda}}
\newcommand{\bOmega}{{\boldsymbol\Omega}}
\newcommand{\bomega}{{\boldsymbol\omega}}

\newcommand{\bmu}{{\boldsymbol\mu}}
\newcommand{\bbeta}{{\boldsymbol\beta}}

\newcommand{\ntb}{\notag\\}

\newcommand{\R}{\mathbb{R}}
\newcommand{\C}{\mathbb{C}}

\newcommand{\I}{\mathbf{I}}

\newcommand{\Lda}{\mathbf{\Lambda}}

\newcommand{\GkH}{\mathbf{G}_{k,u,u}^{H}}
\newcommand{\Gk}{\mathbf{G}_{k,u,u}}

\newcommand{\kt}{{{\widetilde{\mathbf{K}}}_{k,u}}}

\newcommand{\lambdak}{\mathbf{\Lambda}_{k,u}}

\newcommand{\Pc}{P_{\mathrm{c},u}}
\newcommand{\Ps}{P_{\mathrm{s},u}}

\newcommand{\Abar}{{\overline{A}}}

\allowdisplaybreaks

\begin{document}

\title{Energy Efficiency Optimization for Multi-cell Massive MIMO: Centralized and Distributed Power Allocation Algorithms}

\author{
Li~You, Yufei~Huang, Di~Zhang, Zheng~Chang, Wenjin~Wang, and~Xiqi~Gao
	\thanks{
	Part of this work was presented at the IEEE International Conference on Communications in China, Chongqing, China, Aug. 2020 \cite{you20conference}.
	}
	\thanks{	
		Li You, Yufei Huang, Wenjin Wang, and Xiqi Gao are with the National Mobile Communications Research Laboratory, Southeast University, Nanjing 210096, China, and also with the Purple Mountain Laboratories, Nanjing 211100, China (e-mail: liyou@seu.edu.cn; yufei\_huang@seu.edu.cn; wangwj@seu.edu.cn; xqgao@seu.edu.cn).
		
		Di Zhang is with the School of Information Engineering, Zhengzhou University, Zhengzhou 450001, China (e-mail: dr.di.zhang@ieee.org).
		
		Zheng Chang is with the School of Computer Science and Engineering, University of Electronic Science and Technology of China, Chengdu, China, and also with Faculty of Information Technology, University of Jyv\"askyl\"a, FI-40014 Jyv\"askyl\"a, Finland (email: zheng.chang@jyu.fi).
	}
}

\maketitle

\begin{abstract}
This paper investigates the energy efficiency (EE) optimization in downlink multi-cell massive multiple-input multiple-output (MIMO). In our research, the statistical channel state information (CSI) is exploited to reduce the signaling overhead. To maximize the minimum EE among the neighbouring cells, we design the transmit covariance matrices for each base station (BS). Specifically, optimization schemes for this max-min EE problem are developed, in the centralized and distributed ways, respectively. To obtain the transmit covariance matrices, we first find out the closed-form optimal transmit eigenmatrices for the BS in each cell, and convert the original transmit covariance matrices designing problem into a power allocation one. Then, to lower the computational complexity, we utilize an asymptotic approximation expression for the problem objective. Moreover, for the power allocation design, we adopt the minorization maximization method to address the non-convexity of the ergodic rate, and use Dinkelbach's transform to convert the max-min fractional problem into a series of convex optimization subproblems. To tackle the transformed subproblems, we propose a centralized iterative water-filling scheme. For reducing the backhaul burden, we further develop a distributed algorithm for the power allocation problem, which requires limited inter-cell information sharing. Finally, the performance of the proposed algorithms are demonstrated by extensive numerical results.
\end{abstract}



\section{Introduction}

\IEEEPARstart{A}{s} the communication industry develops at an extremely fast speed, we will witness various emerging applications in every aspect of our life, such as agriculture, traffic, health care, and so on \cite{Gupta2015survey5G}. Consequently, the current wireless networks are facing new challenges to fulfill the demands of these new technologies. For example, the form of information shared by people is changing from texts, images to high-definition videos, which brings new requirements on the speed and capacity of wireless communications. In order to cope with this trend, the cellular network needs to be evolved, i.e., from the existing 5G to beyond 5G (B5G) and even 6G \cite{zhang2020mimob5g,Cayamcela2019machinelearing}.
As a key technology in 5G, massive multiple-input multiple-output (MIMO) is maturing and has been widely adopted. It has the merit of bringing about high energy efficiency (EE) and spectral efficiency (SE) gains \cite{Ngo2013MIMO}, and its variant (e.g., cell-free networks, extremely large aperture arrays) will still play an indispensable role in the forthcoming B5G or even 6G eras \cite{Interdonato20cellfree,zhang2020mimob5g}.

In conventional single-cell massive MIMO processing cellular networks, one major obstacle limiting the capacity resides in the inter-cell interference (ICI) resulting from pilot contamination \cite{Huang2012DistriPareto}. The cell-edge user terminals (UTs) tend to undergo severe ICI and thus have poor transmission performance.
As a solution, the base station (BS) cooperation, also known as the coordinated multi-cell precoding, was proposed \cite{Gesbert2010lookinterference}. The main idea of this procedure is that adjacent cells form a cooperative cluster where BSs exchange information with each other.
It can mitigate the co-channel interference and reach a preferable data rate, and has attracted extensive investigations \cite{Gesbert2010lookinterference,Ponukumati2011imperfect,Sun2017BDMA,He2013coorbeam}. In particular, an overview of the multi-cell MIMO cooperation was shown in \cite{Gesbert2010lookinterference}. Additionally, downlink beamforming vectors of a multi-cell network were designed in \cite{Ponukumati2011imperfect} considering two different designing criteria, that is, minimizing the downlink transmit power and maximizing the signal-to-interference-plus-noise ratio. In \cite{Sun2017BDMA} and \cite{He2013coorbeam}, joint power allocation approaches in a coordinated multi-cell downlink system were proposed taking SE and EE as the optimization objectives, respectively.

Note that in downlink multi-cell MIMO,
if the full transmit signals and CSI are shared among cells, the multi-cell network can be equivalently considered as a single cell multi-user system, where the UTs are simultaneously served by a cluster of BSs \cite{Huang2011DistriLimit,Somekh07jointsumrate}.
Nevertheless, this kind of cooperation will result in an enormous backhaul burden because of the vast data sharing. On the contrary, distributed schemes with limited inter-cell information sharing are more favorable, for they require limited backhaul capacity. To this end, in \cite{Ng2008destricooperative}, a distributed beamforming strategy was developed by recasting the downlink beamformer, and the transmit beamforming design was converted to a linear minimum  mean square error estimation problem. A distributed EE-oriented transmission design was raised in \cite{Pan2014totalDistriEE}. By formulating the problem as a non-cooperative game, the research shows that the game reaches a Nash equilibrium and the problem can be tackled in a totally distributed way.

Another critical issue existing in the current massive MIMO system is the energy consumption.
Though the system SE is treated as a more important designing metric in tradition, with the fast growing number of UTs, the power consumption could be remarkably increased.
Due to the ecological and economic considerations, EE-oriented investigations have emerged lately \cite{zappone2015energy,You2020EEoptimal,Tervo2018EE,TTJT17}.
When it comes to energy-efficient precoding design in multi-cell scenario, there are several optimization objectives, such as the global EE, which is denoted as the ratio of all users' sum rate to the total consumed power, or the sum of individual EE of each cell. For instance, energy-efficient power allocation for multi-cell massive MIMO transmission was investigated in \cite{He2013coorbeam}. A coordinated multi-cell multi-user precoding scheme was proposed in \cite{He2014MultiWeightedEE}, aiming to maximize the weighted sum EE.
However, the weighting factors are usually not easy to be adjusted in practice.
Moreover, the fairness cannot be guaranteed when different BSs have different transmit power levels
\cite{Li2016maxminEE}.
Therefore, we tend to adopt the max-min weighted EE as a design criterion to guarantee the minimum EE performance \cite{Du2014distriCOMP,Nguyen2017DistriMISO}.

Most of the aforementioned EE-based transmission approaches rely on perfectly known instantaneous CSI. However, the acquisition error of instantaneous CSI is usually unavoidable in practice, especially in massive MIMO \cite{You15Pilot}. Robust transmission designs exploiting imperfect instantaneous CSI were studied in e.g., \cite{GWCX19,TZQS19}.
On the contrary, the statistical CSI, such as the channel covariance matrix, which reflects the statistical properties of the channels, can be obtained more easily. Moreover, it varies slowly over a much broader time span, which is preferable for power allocation design. Some previous works studied massive MIMO transmission based on statistical CSI \cite{You2018EE,Wang2019EE,You2020Tradeoff,You21EESE}. For instance, in \cite{You2018EE}, an energy-efficient precoding procedure in the beam domain under the multicast massive MIMO scenario was studied. In \cite{Wang2019EE}, the authors extended the EE optimizing question to non-orthogonal unicast and multicast transmission. The resource efficiency optimization transmission strategies were proposed to strike an adaptive EE-SE balance in \cite{You2020Tradeoff,You21EESE}.

Inspired by the above considerations, we aim to investigate the max-min fairness-based EE optimization problem in multi-cell massive MIMO systems exploiting the statistical CSI.
The objective function of our optimization problem is the minimum weighted EE among cells. In general, the considered problem is challenging as the objective is non-convex and exhibits a fractional form.
The major contributions of our work are summarized as follows:
	\begin{itemize}
		\item We first formulate the max-min EE optimization problem to design the transmit covariance matrices at each BS. By exploiting the channel properties in massive MIMO, we obtain the optimal transmit directions at each BS, thus converting the original complex-matrix-valued precoding design problem into a real-vector-valued beam domain power allocation one.
		\item We derive an asymptotic approximation, i.e., the deterministic equivalent (DE), of the ergodic user rate. By doing so, we avoid the complicated expectation calculation in the rate expression. We further handle this optimization problem by solving a sequence of convex optimization subproblems based on the minorization maximization (MM) technique and Dinkelbach's transform.
		\item To address the subproblems, we propose two approaches, i.e., the centralized and distributed ones. For each approach, we put forward generalized water-filling schemes to maximize the Lagrangian function.
		\item Combining all the methods utilized above, we propose low complexity iterative power allocation algorithms with guaranteed convergence for the multi-cell max-min fairness-based EE precoding design. Numerical results illustrate the performance of the proposed algorithms.
	\end{itemize}

The rest of this paper is organized as follows.
In \secref{sec:sysmodel}, we introduce the considered multi-cell multi-user massive MIMO downlink system. The characteristic of the adopted channel model is demonstrated, and the max-min EE problem is formulated.
In \secref{sec:precoding}, the eigenvectors of the transmit power matrices are determined in closed-form, and the MM method and Dinkelbach's transform are applied to address the max-min fractional power allocation design.
In \secref{sec:allocation}, the centralized and distributed algorithms to solve the transformed convex subproblems are described, respectively. We also analyze and compare the complexity of the two approaches. Numerical simulations are conducted in \secref{sec:numerical}. We conclude this paper in \secref{sec:conclusion}.

\emph{Notations:} Throughout this paper, matrices and column vectors are
represented by upper and lower case boldface letters, respectively. We adopt $\mathbb{C}^{M\times N}$ to denote $M\times N$ complex-valued vector space and
$\mathbb{R}^{M\times N}$ to represent $M\times N$ real-valued one.
Moreover, $\mathbf{A}\succeq\mathbf{0}$ indicates a positive semi-definite matrix $\mathbf{A}$, $\mathbf{I}_M$ represents an identity matrix of size $M\times M$. The notation $\diag{\mathbf{x}}$ creates a diagonal matrix with $\mathbf{x}$ along its main diagonal, $\left[ x \right]^+$ represents $\max \{ x,0 \}$.
The superscripts $\left(.\right)^T$, $\left(.\right)^*$ and $\left(.\right)^H$ stand for transpose, conjugate and conjugate-transpose operations, respectively.
The operators are defined by: $\tr{.}$ as the trace operation, $\det\{.\}$ as the determinant operation, $\expect{.}$ as the expectation operation,
$\odot$ as the Hadamard product, and $\triangleq$ for definition, respectively.

\section{System Model and Problem Formulation}\label{sec:sysmodel}
\subsection{System Model}

Consider a $U$-cell massive MIMO downlink system, where each cell $u$ contains one $M$-antenna BS which simultaneously serves $K_u$ multi-antenna UTs. Define the sets for all the cells and UTs as $\mathcal{U} \triangleq \left\{1,\ldots,U\right\}$ and $\mathcal{K} \triangleq \left\{ (k,u) | \,u = 1,\ldots,U,\,k = 1,\ldots,K_u \right\}$, respectively.
Denote the BS in cell $u$ and the $k$th UT in cell $u$ as BS-$u$ and UT-$(k,u)$, respectively.
Each UT-$(k,u)$ has $N_{k,u}$ receive antennas. With $\bx_{k,u}\in \C^{M \times 1}$ and $\bH_{k,u,v} \in \C ^{ N_{k,u} \times M } $ denoting the signal transmitted to UT-$(k,u)$ and the downlink channel matrix from BS-$v$ to UT-$(k,u)$, respectively, the downlink transmission model is expressed as
\begin{align}
{{\bf{y}}_{k,u}} = \bH_{k,u,u}\bx_{k,u} + \sum\limits_{(i,j) \ne (k,u)}{\bH_{k,u,j}\bx_{i,j}} + \bn_{k,u},
\end{align}
where ${{\bf{y}}_{k,u}}\in \C^{N_{k_u} \times 1}$ represents the received signal at UT-$(k,u)$ under an additive Gaussian noise $\bn_{k,u} \in {{\mathbb{C}}^{N_{k_u} \times 1}}$, which is complex circularly symmetric distributed, zero-mean, and has a covariance matrix ${\sigma ^2}{{\mathbf{I}}_{N_{k,u}}}$. The transmitted vector $\bx_{k,u}\left(\forall (k,u) \in \cK\right)$ satisfies $\expect{\bx_{k,u}} = \bzero$, $\expect{\bx_{k,u}\bx^H_{i,j}} = \bzero\left( \forall (i,j)\ne(k,u)\right)$, and its covariance matrix is $\bQ_{k,u} = \expect { \bx_{k,u} \bx_{k,u}^H} \in {\C ^{M \times M}}$.

In our work, we adopt the widely accepted Weichselberger's channel model \cite{Weichselberger06stochastic}. Then, the downlink channel matrix $\bH_{k,u,v}$ is characterized as \cite{Gao09Statistical}
\begin{align}\label{eq:beam_H}
\bH_{k,u,v} = \bU_{k,u,v} \bG_{k,u,v} \bV_{k,u,v}^H,
\end{align}
where ${{\bf{G}}_{k,u,v}} \in {\C ^{N_{k,u} \times M}}$ is a random matrix and referred to as the beam domain channel \cite{You17BDMA}, ${{\bf{U}}_{k,u,v}} \in {\C ^{N_{k,u} \times N_{k,u}}}$ and ${{\bf{V}}_{k,u,v}} \in {\C ^{M\times M}}$ are deterministic unitary matrices, denoting the eigenvector matrices of the UT and the BS correlation matrix, respectively. Note that the entries in $\bG_{k,u,v}$ are statistically uncorrelated \cite{Gao09Statistical}.
The structure of the downlink MIMO channel model shown in \eqref{eq:beam_H} describes the downlink channel spatial correlations.
It is proved in \cite{You15Pilot} that, as $M \to \infty $, $\bV_{k,u,v}$ becomes asymptotically identical, i.e.,
\begin{align}\label{eq:V identical}
\bV_{k,u,v}  \mathop = \limits^{M \to \infty } \bV,
\end{align}
where ${\bf{V}}$ only relates to the BS antenna array topologies and is irrelevant to the locations of UTs.
For instance, in the case of uniform linear array (ULA) antenna configurations employed at the BS, the discrete Fourier transform (DFT) matrix is a good approximation of $\bV$ \cite{You15Pilot,You16Channel}. Note that for practical cases where $M$ is large and finite, $\bV$ can be seen as a good approximation to $\bV_{k,u,v}$ \cite{Adhikary13Joint}.

In this work, we design the precoding problem based on the statistical CSI instead of instantaneous CSI. Specifically, we adopt the concept of eigenmode channel coupling matrix in \cite{Gao09Statistical}, which is defined as
\begin{align}
\bOmega_{k,u,v} = \expect { {{\bf{G}}_{k,u,v}} \odot {\bf{G}}_{k,u,v}^* }  \in { \R ^{N_{k,u} \times M}}.
\end{align}
Throughout the paper, we assume that only the statistical CSI $\bOmega_{k,u,v}$ is accessible to each BS-$v$ while each UT-$(k,u)$ has access to its own instantaneous CSI through proper pilot design \cite{Sun2015Beam}.

\subsection{Problem Formulation}
We treat $\bn_{k,u}^{\prime} = \sum\nolimits_{(i,j) \ne (k,u)} { \bH_{k,u,j} \bx_{i,j} }  + {{\bf{n}}_{k,u}}$ as the aggregate interference-plus-noise, whose covariance matrix $\bK_{k,u}\in { \C ^{N_{k,u} \times N_{k,u}}}$ is also accessible at each UT-$(k,u)$. For a worst-case design\footnote{Note that the worst-case design means that for a fixed noise covariance, the worst case is that the noise follows a Gaussian distribution
in the sense that the corresponding rate is a lower bound of the exact
rate \cite{Hassibi2003How}.}, $\bn_{k,u}^{\prime}$ is supposed to be a Gaussian noise with covariance \cite{Hassibi2003How}
\begin{align}\label{eq:K_definition}
\bK_{k,u} = {\sigma ^2} \bI_{N_{k,u}} + \sum\limits_{(i,j) \ne (k,u) } {\expect { \bH_{k,u,j} \bQ_{i,j} {{\bf{H}}_{k,u,j}^H} } }.
\end{align}

Under the above assumptions, we calculate the ergodic data rate of UT-$(k,u)$ as in \cite{Lu2019Robust,wu18beam}, i.e.,
\begin{align}\label{eq:ergodic_rate}
{R_{k,u}} & = \expect { \log \det \left( \bK_{k,u} + \bH_{k,u,u} \bQ_{k,u} \bH_{k,u,u}^H \right) } \ntb
  &  \quad \quad \qquad - \log\det\left( \bK_{k,u} \right)\ntb
& =  \expect { \log\det\left({\kt} + \bG_{k,u,u} \bV^H \bQ_{k,u} \bV \bG_{k,u,u}^H\right) } \ntb
 & \quad \quad \qquad - \log\det\left({\kt}\right),
\end{align}
where the second equality can be obtained through first rewriting $\bH_{k,u,u} $ by \eqref{eq:beam_H} and \eqref{eq:V identical} as $\bH_{k,u,u} = \bU_{k,u,v} \bG_{k,u,v} \bV^H$ and further exploiting Sylvester's determinant identity, i.e., $\det ({\bf{I}} + {\bf{WZ}}) = \det ({\bf{I}} + {\bf{ZW}})$. Moreover, $\kt\in { \C ^{N_{k,u} \times N_{k,u}}}$ in \eqref{eq:ergodic_rate} is expressed as
\begin{align}\label{eq:K_tilde}
{\kt} &\triangleq \bU_{k,u,u}^H \bK_{k,u} \bU_{k,u,u} \ntb
& = {\sigma ^2} \bI_{N_{k,u}} + \sum\limits_{(i,j) \ne (k,u)} {\underbrace {\expect { \bG_{k,u,j} \bV^H \bQ_{i,j} \bV \bG_{k,u,j}^H} }_{ \triangleq {{\bf{\Pi}}_{k,u,j}}({{\bf{V}}^H}{{\bf{Q}}_{i,j}}{\bf{V}})}}.
\end{align}
The matrix-valued function ${{\bf{\Pi}}_{k,u,j}}({\bf{X}})$ defined in \eqref{eq:K_tilde} can be proved diagonal by applying the uncorrelated properties of the elements in $\bG_{k,u,v}$. Moreover, the $n$th diagonal element of ${{\bf{\Pi}}_{k,u,j}}({\bf{X}})$ can be calculated as
\begin{align}\label{eq:piX}
{\left[ {{{\bf{\Pi}}_{k,u,j}}({\bf{X}})} \right]_{n,n}}
= \tr { \diag{\left( {\left[\bOmega_{k,u,j} \right]}_{n,:}\right)^T}  \bX }.
\end{align}

In this context, the EE of each cell $u$ is defined as
\begin{align}\label{eq:SE_definition}
\mathrm{EE}_{u} &= \frac{ W R_{u}}{P_{u}} \ntb
&= \frac{W \sum\nolimits_{k = 1}^{K_u}{ R_{k,u} }}{\xi_u \sum\nolimits_{k = 1}^{K_u}{ \tr { \bQ_{k,u} } } + M\Pc + \Ps},\; \forall u\in\cU,
\end{align}
where $W$ represents the communication bandwidth, $R_{u}$ is the sum rate of cell $u$ and $P_{u}$ is the power consumption required to operate cell $u$.
Note that this power consumption model is widely adopted \cite{Ng12Energy,zappone2015energy}. Specifically, the constants $\xi_u(>1)$, $\Ps$, and $\Pc$ are the inverse of the power amplifier efficiency, the dynamic power consumed per BS antenna and the basic power consumption of BS-$u$, respectively, and $\tr{ \bQ_{k,u}}$ denotes the transmit power of UT-$k$ in BS-$u$.

The main focus of our work is to investigate the multi-cell fairness-based EE problem. To this end, we maximize the minimum weighted EE among all cells, which is characterized as
\begin{align}\label{eq:EE maximization problem0}
\mathcal{P}:\;\underset{\bQ_{k,u},\forall (k,u)} \max \quad & \underset{u} \min \ \left\{ w_u \mathrm{EE}_u \right\} \ntb
{\mathrm{s.t.}}\quad
& \sum\limits_{k=1}^{K_u}{\tr{ \bQ_{k,u} }} \le P_{\max,u},\; \forall u \in \cU \ntb
& \bQ_{k,u} \succeq \bzero,\; \forall (k,u)\in \mathcal{K},
\end{align}
where the weighting factor $w_u$ indicates the priority of cell $u$ and $P_{\max,u}$ is the maximum value of the transmit power of BS-$u$.

\section{Energy Efficient Precoding Design}\label{sec:precoding}
In this section, we investigate the precoding design for the minimum weighted EE maximization problem $\mathcal{P}$ in \eqref{eq:EE maximization problem0}. Directly handling $\cP$ is of high complexity, since $\bQ_{k,u},\forall (k,u)$ is a high dimensional complex-valued matrix. Therefore, we start by decomposing the transmit covariance matrix as $\bQ_{k,u} = \bPsi_{k,u}\bLambda_{k,u}\bPsi_{k,u}^H$ based on the eigenvalue decomposition. By doing so, we derive the transmit directions of the signals, which are represented by the columns of $\bPsi_{k,u}$.
In addition, the power allocated to each transmit direction is represented by the diagonal elements of $\bLambda_{k,u}$, i.e., the eigenvalues of $\bQ_{k,u}$.

\subsection{Optimal Transmit Direction}
For the transmit eigenmatrix $\bPsi_{k,u}$, we apply the similar technique adopted in \cite{You2020EEoptimal,You2018EE}, and obtain the following proposition.
\begin{prop}\label{prop1}
The optimal transmit eigenmatrix $\bPsi_{k,u}$ for problem $\cP$ in \eqref{eq:EE maximization problem0} is given as $\bPsi_{k,u}^{\mathrm{opt}} = \bV$, where $\bV$ is the transmit correlation matrix of the channel, and the transmit covariance matrices become
	\begin{equation}
	\bQ_{k,u}^{\mathrm{opt}} = \bV\bLambda_{k,u}\bV^H,\; \forall (k,u) \in \cK.
	\end{equation}
\end{prop}

From \emph{Proposition \ref{prop1}}, we can obtain the optimal transmit eigenmatrix $\bPsi_{k,u}$ in closed-form for problem $\cP$ in \eqref{eq:EE maximization problem0}. Therefore, we can reduce the corresponding optimization complexity by converting the problem $\cP$ into a beam domain power allocation problem. With a slight abuse of notations, we define a matrix set $\Lda \triangleq \left\{ {{\Lda}_{1,1}},\ldots ,{{\Lda}_{K_U,U}} \right\}$, and the corresponding power allocation problem is formulated as
\begin{align}\label{eq:problembeam}
\mathcal{P}_1:\;\underset{\mathbf{\Lambda}} \max \quad & \underset{u}\min \ \left\{ \frac{w_u W \sum\nolimits_{k=1}^{K_u}R_{k,u}\left(\Lda\right)}{ \xi_u \sum\nolimits_{k = 1}^{K_u}{ \tr { \bLambda_{k,u} } } + M\Pc + \Ps} \right\} \ntb
{\mathrm{s.t.}}\quad
& \sum\limits_{k=1}^{K_u}{\tr{ \bLambda_{k,u} }} \le P_{\max,u},\; \forall u \in \cU \ntb
& \bLambda_{k,u} \succeq \bzero,\;\bLambda_{k,u} \;\mathrm{diagonal},\;\forall (k,u)\in \mathcal{K},
\end{align}
where
\begin{align}\label{eq:kku}
R_{k,u}\left(\Lda\right)  =  &  \underbrace{\expect { \log \det \left( \kt\left(\Lda\right)+ \Gk\lambdak\GkH \right)  }}_{\triangleq A_{k,u}\left( \Lda \right)} \ntb
&- \underbrace{ \log \det \left( \kt\left(\Lda\right) \right) }_{\triangleq B_{k,u}\left( \Lda \right)}  .
\end{align}
Note that $\kt\left(\Lda\right)$ in \eqref{eq:kku} is a matrix-valued function defined as
\begin{align}
\kt\left(\Lda\right) \triangleq {\sigma ^2} \I_{N_{k,u}} +  \sum\limits_{(i,j)\ne (k,u)}{ {{\bf{\Pi}}_{k,u,j}}\left(\bLambda_{i,j}\right)},
\end{align}
and the definition of ${{\bf{\Pi}}_{k,u,j}}({\bf{X}})$ is given in \eqref{eq:piX}.

Note that the number of variables in problem $\cP_1$, compared with problem $\cP$, is reduced from $M^2\times\left(\sum_{u=1}^{U}Ku\right)$ to $M\times\left(\sum_{u=1}^{U}Ku\right)$. This reduction is especially significant for a massive MIMO system with large $M$.

\subsection{Power Allocation Design}
Next, we focus on the optimization strategies for problem $\cP_1$. Note that the objective of $\cP_1$ is the minimum of several fractional functions, where the denominator of each fractional function is linear over $\bLambda$ and the numerator exhibits a non-concave form.
In line with this, we put forward a power allocation strategy to maximize the minimum EE by means of fractional programming and iterative optimization.

Firstly, we can see that for each UT-$(k,u)$, direct calculation of the ergodic rate, $R_{k,u}\left(\bLambda\right)$, which involves the expectation operation on large-dimensional matrices, would incur high computational complexity.
To obtain a tractable approximation for the ergodic rate, we apply the large-dimensional matrix theory \cite{Lu16Free,Couillet11Random} and replace the ergodic rate with its DE expression. The main idea of DE is to introduce several auxiliary variables to iteratively calculate the object function, avoiding the high-dimensional expectation operation. Note that the DE expression is proved to be an asymptotically accurate approximation of the objective and especially suitable in massive MIMO systems \cite{Wen11On}.
Specially,
the DE of $A_{k,u}\left( \Lda \right)$ in \eqref{eq:kku} is computed by
\begin{align}\label{eq:DE}
\Abar_{k,u} \left( \Lda \right) = & \log \det \left( \mathbf{I}_{M}+\mathbf{\Gamma }_{k,u}{{\mathbf{\Lambda }}_{k,u}} \right) \ntb
+\log &\det \left( \mathbf{\widetilde{\Gamma }}_{k,u}+\kt\left( \Lda \right) \right) -\tr { \mathbf{I}_{N_{k,u}}-{{\mathbf{\widetilde{\Phi }}}^{-1}_{k,u}} },
\end{align}
where the auxiliary variables are represented as
\begin{align}
\label{eq:gamma_1}
\mathbf{\Gamma }_{k,u}&={\bT_{k,u}}\left( {{{\mathbf{\widetilde{\Phi }}}^{-1}_{k,u}}}{ {\left(\kt\left( \Lda \right)\right)} ^{-1}} \right),\\
\label{eq:gamma_2}
\mathbf{\widetilde{\Gamma }}_{k,u}&={\bF_{k,u}}\left( { {{{\mathbf{\Phi }}^{-1}_{k,u}}}}{{ {\mathbf{\Lambda }}_{k,u}} } \right),\\
\label{eq:phi_1}
\mathbf{\widetilde{\Phi }}_{k,u}&=\mathbf{I}_{N_{k,u}} + \mathbf{\widetilde{\Gamma }}_{k,u} {{\left(\kt\left( \Lda \right)\right)}^{-1}},\\
\label{eq:phi_2}
\mathbf{\Phi }_{k,u}&=\mathbf{I}_{M} + \mathbf{\Gamma }_{k,u} {{\mathbf{\Lambda }}_{k,u}}.
\end{align}
The matrix-valued functions $\bT_{k,u} (\bX) $ and $\bF_{k,u} (\bY) $ are defined by ${\bT_{k,u}}  \left( \bX \right) \triangleq \expect { \bG_{k,u,u}^H \bX \Gk }\in \C^{M\times M}$ and ${\bF_{k,u}}  \left( \bY \right) \triangleq \expect { \Gk \bY \bG_{k,u,u}^H }\in \C^{N_{k,u}\times N_{k,u}}$, respectively. They are both diagonal functions since the elements of $\Gk$ are zero-mean and statistically uncorrelated, and the corresponding diagonal elements are given by
\begin{align}\label{eq:T_element}
&{\left[ \bT_{k,u}(\bX) \right]_{m,m}} = \tr { \diag{{\left[\bOmega_{k,u,u} \right]}_{:,m}} \bX },\\
\label{eq:F_element}
&{\left[ \bF_{k,u}(\bY) \right]_{n,n}} = \tr { \diag{ \left(\left[\bOmega_{k,u,u} \right]_{n,:}\right)^T} \bY },
\end{align}
respectively.
Then, with the aid of $\Abar_{k,u} \left( \Lda \right)$, we turn to consider the following problem
\begin{align}\label{eq:problemDE1}
\mathcal{P}_2:\; \underset{\mathbf{\Lambda}} \max \quad & \underset{u} \min \ \left\{ \frac{w_u W \sum\nolimits_{k = 1}^{K_u} { \left\{ \Abar_{k,u}\left( \Lda \right)- B_{k,u}\left( \Lda \right) \right\}} }{  \xi_u \sum\nolimits_{k = 1}^{K_u}{ \tr { \bLambda_{k,u} } } + M\Pc + \Ps  }\right\} \ntb
{\mathrm{s.t.}}\quad
& \sum\limits_{k=1}^{K_u}{\tr{ \bLambda_{k,u} }} \le P_{\max,u},\; \forall u \in \cU \ntb
& \bLambda_{k,u} \succeq \bzero,\; \bLambda_{k,u} \;\mathrm{diagonal},\; \forall (k,u)\in \mathcal{K}.
\end{align}

For iterative convex optimization methods, we resort to the MM procedure \cite{Sun2017MM}. The main idea of MM procedure is to convert the original non-convex program into a series of solvable subproblems. Following the MM procedure, in each subproblem, we find a surrogate function to approximate the ergodic rate. Specially, we replace $B_{k,u}\left( \Lda \right)$ in \eqref{eq:kku}, the negative term of the rate,  with its first-order Taylor expansion. Then, the numerator becomes a concave function of $\bLambda$.
Therefore, problem $\mathcal{P}_2$ is solved through iteratively solving the subproblems as follows
\begin{align}\label{eq:problem MM}
\mathcal{P}_3^{(\ell)}:\; &\Lda^{(\ell+1)} =  \ntb \underset{\mathbf{\Lambda}} {\arg\max} \quad & \underset{u} \min \ \left\{  \frac{w_u W \sum\nolimits_{k = 1}^{K_u} { \left\{ \Abar_{k,u}\left( \Lda \right)- \triangle B_{k,u}^{(\ell)}\left( \Lda \right) \right\}} }{  \xi_u \sum\nolimits_{k=1}^{K_u}{ \tr { \bLambda_{k,u} } } + M\Pc + \Ps  } \right\}\ntb
{\mathrm{s.t.}}\quad
& \sum\limits_{k=1}^{K_u}{\tr{ \bLambda_{k,u} }} \le P_{\max,u},\; \forall u \in \cU \ntb
& \bLambda_{k,u} \succeq \bzero,\; \bLambda_{k,u} \;\mathrm{diagonal},\; \forall (k,u)\in \mathcal{K},
\end{align}
where $\Lda^{\left( \ell \right)} \triangleq \left\{ {{\Lda}^{(\ell)}_{1,1}},\ldots ,{{\Lda}^{(\ell)}_{K_U,U}} \right\}$ and $\ell$ denotes the iteration index. The first-order Taylor expansion term is expressed as
\begin{align}
\triangle B_{k,u}^{(\ell)} \left(\Lda\right) =  B_{k,u}&\left(\Lda^{(\ell)}\right) \ntb
+ \sum\limits_{ (i,j) \ne (k,u) } &{ \tr{ \left( \frac{\partial  B_{k,u}\left(\Lda^{(\ell)}\right) }{ \partial \bLambda_{i,j} } \right)^T \left( \bLambda_{i,j} -\bLambda^{(\ell)}_{i,j} \right) } }.
\end{align}
Moreover, the derivative $\frac{\partial  B_{k,u}\left(\Lda^{(\ell)}\right) }{ \partial \bLambda_{i,j} }$ is derived as
\begin{align}\label{eq:derivative_CCCP}
\bD_{k,u,j}^{(\ell)} &= \frac{\partial  B_{k,u}\left(\Lda^{(\ell)}\right) }{ \partial \bLambda_{i,j} } \ntb
&= \sum\limits_{n = 1}^{N_{k,u}} {\frac{ {\widehat {\bR} }_{k,u,j,n} }{ {\sigma}^2 + \tr { \Lda^{(\ell)}_{\backslash (k,u)} {\widehat {\bf R}}_{k,u,q,n}}}},\quad (i,j) \ne (k,u),
\end{align}
where $\bLambda^{(\ell)}_{\backslash (k,u)}= \sum\nolimits_{(p,q) \ne (k,u)} {\bLambda^{(\ell)}_{p,q}}$ and ${{\widehat {\bf R}}_{k,u,j,n}} = \diag{ {\bomega } _{k,u,j,n}}$ with ${{\bomega } _{k,u,j,n}^{T}}$ being the $n$th row of ${{\bf{\Omega }}_{k,u,j}}$. Note that $\bD_{k,u,j}^{(\ell)}$ is a diagonal matrix with its $t$th diagonal entry given by
\begin{align}
&\left[\bD_{k,u,j}^{(\ell)}\right]_{t,t} \ntb
&\qquad= \sum\limits_{n = 1}^{N_{k,u}} { \frac{ \left[\bOmega_{k,u,j}\right]_{n,t} }{ {\sigma}^2 +  \sum\limits_{(p,q) \ne (k,u)} {\sum\limits_{m = 1}^{M} { \left[\bOmega_{k,u,q}\right]_{n,m} \left[ \bLambda^{(\ell)}_{p,q} \right]_{m,m} } } } }  .
\end{align}

Dinkelbach's transform can obtain the global optimal solution of a max-min concave-linear fractional programming by solving a sequence of convex problems \cite{zappone2015energy,shen2018fractional}. Specifically, for each max-min fractional problem $\cP_3^{(\ell)}$ in \eqref{eq:problem MM}, it can be equivalently handled by solving a series of concave maximization subproblems as follows
\begin{align}\label{eq:problemDinkel}
\mathcal{P}_4^{(\ell),[t]}:\; &\bLambda^{(\ell),[t+1]} = \ntb \underset{\bLambda} {\arg\max} \quad & \underset{u} \min \  \left\{ w_u W \overline{R}^{(\ell)}_u\left(\bLambda \right)- \eta^{(\ell),[t]}  P_u\left(\bLambda\right)  \right\}\ntb
{\mathrm{s.t.}}\quad
& \sum\limits_{k=1}^{K_u}{\tr{ \bLambda_{k,u} }} \le P_{\max,u},\; \forall u \in \cU \ntb
& \bLambda_{k,u} \succeq \bzero,\; \bLambda_{k,u} \;\mathrm{diagonal},\; \forall (k,u)\in \mathcal{K},
\end{align}
where $\Lda^{ (\ell),[t+1]} \triangleq \left\{ { \Lda^{(\ell),[t+1]}_{1,1},\ldots ,\Lda^{(\ell),[t+1]}_{K_U,U}} \right\}$, $t$ is the iteration index of Dinkelbach's transform and
\begin{align}\label{eq:R ell u}
\overline{R}^{(\ell)}_u \left(\bLambda \right) & = \sum\limits_{k = 1}^{K_u} { \left\{ \Abar_{k,u}\left( \Lda \right)- \triangle B_{k,u}^{(\ell)}\left( \Lda \right) \right\}}, \\
\label{eq:power}
P_u\left(\bLambda\right) & = \xi_u \sum\limits_{k=1}^{K_u}{ \tr { \bLambda_{k,u} } } + M\Pc + \Ps, \\
\label{eq:eta update}
\eta^{(\ell),[t]} & = \underset{u}{\min} \left\{ \eta^{(\ell),[t]}_u \right\} = \underset{u}{\min} \left\{\frac{ w_u W \overline{R}^{(\ell)}_u\left(\bLambda^{(\ell),[t]} \right)}{P_u\left(\bLambda^{(\ell),[t]}\right) }\right\}.
\end{align}
Particularly, we summarize Dinkelbach's transform-based max-min EE maximization power allocation procedure in \alref{algMMDinkel}.

\begin{algorithm}[htbp]
	\caption{Max-Min EE Power Allocation Algorithm}
	\label{algMMDinkel}
	\setstretch{1.2}
	\begin{algorithmic}[1]
		\Require Initial power allocation matrix $\bm{\Lambda}^{\left(0\right)}$, channel statistics $\bOmega_{k,u,v}$, iteration thresholds $\epsilon_1$, $\epsilon_2$.
		\Ensure Power allocation matrix $\bm{\Lambda}$.
		\State Initialization: $\ell=0$,  $\underset{u}{\min} \left\lbrace w_u \overline{\mathrm{EE}}^{(-1)}_u \right\rbrace = 0$.
		\State For $\forall u \in \cU$, calculate
		\begin{equation}\label{eq:EE}
		\overline{\mathrm{EE}}^{\left(\ell\right)}_u=\frac{ W \sum\limits_{k = 1}^{K_u} { \left\{ \Abar_{k,u}\left( \Lda ^{\left(\ell\right)}\right)- \triangle B_{k,u}^{(\ell)}\left( \Lda^{\left(\ell\right)} \right) \right\}} }{P_u\left(\bm{\Lambda}^{\left(\ell\right)}\right)}.
		\end{equation}
		\While{$\left| \underset{u}{\min} \left\lbrace 		w_u\overline{\mathrm{EE}}^{\left(\ell\right)}\right\rbrace-\underset{u}{\min} \left\lbrace w_u\overline{\mathrm{EE}}^{\left(\ell-1\right)}\right\rbrace\right|\geq\epsilon_1$}
		\State Initialization: $t=0$, $ \bm{\Lambda}^{(\ell),[-1]}=\bm{\Lambda}^{\left(\ell\right)} $, $\eta^{(\ell),[-1]}=0$.
		\While{$\left|\eta^{(\ell),[t]}-\eta^{(\ell),[t-1]}\right| \geq\epsilon_2$}
		\State Set $t = t+1$.
		\State Solve problem $\cP_4^{(\ell),[t-1]}$ in \eqref{eq:problemDinkel} with $\eta^{(\ell),[t-1]}$ and obtain the solution $\bm{\Lambda}^{(\ell),[t]}$.
		\State Update $ \eta^{(\ell),[t]} $ with \eqref{eq:eta update}.
		\EndWhile
		\State Set $\ell = \ell + 1$.
		\State Set $\bm{\Lambda}^{(\ell)} = \bm{\Lambda}^{(\ell-1),[t]}$.
		\State Update $\overline{\mathrm{EE}}_u^{\left(\ell\right)}$ with \eqref{eq:EE}.
		\EndWhile
		\State \Return{$\bm{\Lambda}=\bm{\Lambda}^{\left(\ell\right)}$}.
	\end{algorithmic}
\end{algorithm}

\section{Energy-Efficient Power Allocation Algorithm}\label{sec:allocation}
It is worth noting that each subproblem $\cP_4^{(\ell),[t]}$ in \eqref{eq:problemDinkel} is convex and can be solved via utilizing the classical convex optimization methods \cite{Boyd04Convex}. However, directly solving it might lead to high computational complexity, especially in the massive MIMO scenario.
We therefore focus on developing low-complexity algorithms for $\cP_4^{(\ell),[t]}$ in this section.

Generally, power allocation design in the multi-cell scenario is conducted in a centralized way, where the BSs jointly design the power allocation matrices to suppress the ICI and reach a favorable performance. The centralized approach is feasible in many scenarios. However, in massive MIMO transmissions, the exchange of CSI between BSs might cause large backhaul burdens. On the contrary, a distributed implementation of the power allocation scheme requiring only a few parameters is preferable to lighten the burden of the backhaul links.
The distributed approach offers a trade-off between performance gains and the amount of overhead in the backhaul and feedback channels. In the following, we first introduce a centralized joint power allocation approach with full CSI sharing. Then, we propose a distributed scheme with limited inter-cell cooperation to solve the sub-optimal problem in \eqref{eq:problemDinkel}.

\subsection{Centralized Power Allocation}
We first focus on the centralized algorithm for problem $\mathcal{P}_4^{(\ell),[t]}$ in \eqref{eq:problemDinkel}.
Noting that the objective function of problem $\mathcal{P}_4^{(\ell),[t]}$ is not continuously differentiable due to the minimum operation, 
we introduce a new variable $z$ to represent the minimum weighted EE and arrive at an equivalent formulation of \eqref{eq:problemDinkel} as follows
\begin{subequations}\label{eq:problemz}
	\begin{align}\label{eq:problemzobjective}
	\mathcal{P}_5^{(\ell),[t]}:\; \underset{\bLambda, \ z} \max \quad & z \\
	{\mathrm{s.t.}}\quad\label{eq:problemzconstraint1}
	& z \le \ w_u \left\{   W \overline{R}^{(\ell)}_u\left(\bLambda \right)- \eta^{(\ell),[t]}  P_u\left(\bLambda\right)  \right\},\ntb
	&\qquad \qquad \qquad \qquad \qquad \forall u \in \cU \\
	\label{eq:problemzconstraint2}
	& \sum\limits_{k=1}^{K_u}{\tr{ \bLambda_{k,u} }} \le P_{\max,u},\; \forall u \in \cU \\
	& \bLambda_{k,u} \succeq \bzero,\; \bLambda_{k,u} \;\mathrm{diagonal},\; \forall (k,u)\in \mathcal{K}.
	\end{align}
\end{subequations}

In the following, we apply the dual method \cite{Li2020PhysicalMulti} to solve problem \eqref{eq:problemz}. Specifically, we calculate the dual function, i.e., the maximum of the Lagrangian while fixing the dual variables. Then, we apply the subgradient method to update the dual variables. The diminishing stepsize rule guarantees the convergence of the subgradient method \cite{Boyd04Convex}.
Regarding the solution to \eqref{eq:problemz}, we have the following proposition.

\begin{prop}
Denote the optimal power allocation matrices of problem $\cP_5^{(\ell),\left[t\right]}$ in \eqref{eq:problemz} as $\bLambda^{(\ell),[t+1]}_{k,u},\;\forall (k,u)\in\cK$. Then, the $m$th diagonal element of $\bLambda^{(\ell),[t+1]}_{k,u}$, i.e., $\lambda^{(\ell),[t+1]}_{k,u,m}$, satisfies \eqref{eq:newtoncentral}, shown at the top of next page,
\begin{figure*}
	\begin{equation}\label{eq:newtoncentral}
	\left\{ \begin{array}{l}
	\frac{ \beta_{u} w_u W\gamma_{k,u,m}}{{1 + \gamma _{k,u,m}\lambda^{(\ell),[t+1]}_{k,u,m}}}
	+ \sum\limits_{a \ne k }^{K_u} {\sum\limits_{n = 1}^{N_{a,u}}  \frac{\beta_u w_u  W \left[\bOmega_{a,u,u}\right]_{n,m}} {\widetilde \gamma _{a,u,n} + \left[ \widetilde{\mathbf{K}}_{a,u} (\bLambda) \right]_{n,n} }  }
	+ \sum\limits_{j \ne u }^{U} {\sum\limits_{a = 1 }^{K_j} {\sum\limits_{n = 1}^{N_{a,j}} { \frac{ \beta_j w_j  W \left[\bOmega_{a,j,u}\right]_{n,m}} {\widetilde \gamma _{a,j,n} + \left[ \widetilde{\mathbf{K}}_{a,j} (\bLambda) \right]_{n,n} } } }} = \eta^{(\ell),[t]}\xi_u + \mu_u^{[t+1]}\\
	\quad \quad \quad\quad \quad \quad\quad \quad + \beta_{u} w_u W \sum\limits_{a \ne k}^{K_u}{d^{(\ell)}_{a,u,u,m}} + \sum\limits_{j \ne u }^{U} {\beta_j w_j  W \sum\limits_{a = 1}^{K_j} d^{(\ell)}_{a,j,u,m} }, \quad \quad\quad
	\mu_u^{[t+1]} < \chi_{k,m}^{(\ell),[t+1]}\\
	\lambda^{(\ell),[t+1]}_{k,u,m} = 0,\quad\quad\quad\quad\quad\quad\quad\quad\quad\quad\quad\quad\quad\quad\quad\quad\quad \quad\quad\quad\quad\quad\quad\quad\;\;\; \mu_u^{[t+1]} \geq \chi_{k,m}^{(\ell),[t+1]}
	\end{array} \right.
	\end{equation}
\end{figure*}
where $d^{(\ell)}_{a,u,u,m}$ and $\gamma_{k,u,m}$ stand for the $m$th diagonal element of $\bD^{(\ell)}_{a,u,u}$ and $\boldsymbol{\Gamma}_{k,u}$, respectively.
Moreover, the auxiliary variables $\chi_{k,m}^{(\ell),[t+1]}$ and $\cT_{k, m, a,u}$ are written as \eqref{eq:centraaux} and \eqref{eq:centraauxt} shown at the top of next page, respectively.
\begin{figure*}
	\begin{align}
	\chi_{k,m}^{(\ell),[t+1]} = &\gamma_{k,u,m} - \eta^{(\ell),[t]}\xi_u -  \beta_{u} w_u W \sum\limits_{a \ne k}^{K_u}{d^{(\ell)}_{a,u,u,m}} + \sum\limits_{a \ne k }^{K_u} {\sum\limits_{n = 1}^{N_{a,u}} { \frac{\beta_u w_u  W \left[\bOmega_{a,u,u}\right]_{n,m}} {\widetilde \gamma _{a,u,n} + {\sigma}^2 + \sum\nolimits_{\left(a^{\prime},q,m^{\prime}\right) \in \cT_{k, m, a,u}} \lambda_{a^{\prime},q,m^{\prime}}^{(\ell),[t+1]} \left[\bOmega_{a,u,u}\right]_{n,m^{\prime}} }} } \ntb
	& -\sum\limits_{j \ne u }^{U} {\beta_j w_j  W \sum\limits_{a = 1}^{K_j} d^{(\ell)}_{a,j,u,m} }  + \sum\limits_{j \ne u }^{U} {\sum\limits_{a = 1 }^{K_j} {\sum\limits_{n = 1}^{N_{a,j}} { \frac{ \beta_j w_j  W \left[\bOmega_{a,j,u}\right]_{n,m}} {\widetilde \gamma _{a,j,n} + {\sigma}^2 +  \sum\nolimits_{\left(a^{\prime},q,m^{\prime}\right) \in \cT_{k, m, a,j}} \lambda_{a^{\prime},q,m^{\prime}}^{(\ell),[t+1]} \left[\bOmega_{a,j,u}\right]_{n,m^{\prime}}} } }}\label{eq:centraaux}\\
	\cT_{k, m, a,u} =&\left\{\left(a^{\prime},q,m^{\prime}\right) | \left(a^{\prime},q\right) \neq (a,u),\left(a^{\prime}, q,m^{\prime}\right) \neq(k,u, m), \left(a^{\prime},q,m^{\prime}\right) \in\cK, m^{\prime} \in\{1, \ldots, M\}\right\}\label{eq:centraauxt}
	\end{align}
\end{figure*}
\end{prop}
\begin{proof}
	See \appref{B}.
\end{proof}

The solutions in \eqref{eq:newtoncentral} suggest that the optimal power allocation matrices follow a water-filling structure. We summarize the centralized iterative generalized water-filling power allocation algorithm in \alref{algCentral}. In the multi-cell multi-user scenario, it is hard to find closed-form solutions to equation \eqref{eq:newtoncentral}. In order to solve this problem, we utilize Newton's method \cite{Burden14numerical}, where the auxiliary functions $\varrho_{k,u,m}^{(\ell),[t]}\left(x_{k,u,m}\right)$ and $\varrho_{k,u,m}^{\prime(\ell),[t]} \left(x_{k,u,m}\right)$ in Step $8$ of \alref{algCentral} are defined as \eqref{varrho1} and \eqref{varrho2} shown at the top of next page, respectively.
\begin{figure*}
	\begin{align}\label{varrho1}
	\varrho_{k,u,m}^{(\ell),[t]}\left(x_{k,u,m}\right) =& \frac{ \beta^{[t]}_{u} w_u W\gamma_{k,u,m}}{{1 + \gamma _{k,u,m}x_{k,u,m}}} - \beta^{[t]}_{u} w_u W \sum\limits_{a \ne k}^{K_u}{d^{(\ell)}_{a,u,u,m}} - \eta^{(\ell),[t]}\xi_u - \mu_u^{[t]} - \sum\limits_{j \ne u }^{U} {\beta^{[t]}_j w_j  W \sum\limits_{a = 1}^{K_j} d^{(\ell)}_{a,j,u,m} }\ntb
	&+ \sum\limits_{a \ne k }^{K_u} {\sum\limits_{n = 1}^{N_{a,u}} { \frac{\beta^{[t]}_u w_u  W \left[\bOmega_{a,u,u}\right]_{n,m}} {\widetilde \gamma _{a,u,n} + {\sigma}^2 +x_{k,u,m}\left[\bOmega_{a,u,u}\right]_{n,m}+\underset{\in \cT_{k, m, a,u}}{\underset{\left(a^{\prime},q,m^{\prime}\right) }{\sum}}
				x_{a^{\prime},q,m^{\prime}} \left[\bOmega_{a,u,u}\right]_{n,m^{\prime}} }} } \ntb
	& + \sum\limits_{j \ne u }^{U} {\sum\limits_{a = 1 }^{K_j} {\sum\limits_{n = 1}^{N_{a,j}} { \frac{ \beta^{[t]}_j w_j  W \left[\bOmega_{a,j,u}\right]_{n,m}} {\widetilde \gamma _{a,j,n} + {\sigma}^2 + x_{k,u,m}\left[\bOmega_{a,j,u}\right]_{n,m} + \underset{\in \cT_{k, m, a,j}}{\underset{{\left(a^{\prime},q,m^{\prime}\right) }}{\sum} }
					x_{a^{\prime},q,m^{\prime}} \left[\bOmega_{a,j,u}\right]_{n,m^{\prime}}} } }}
	\end{align}
	\begin{align}\label{varrho2}
	\varrho_{k,u,m}^{\prime(\ell),[t]} \left(x_{k,u,m}\right) =& -\frac{ \beta^{[t]}_{u}w_u W\left(\gamma _{k,u,m}\right)^2}{\left(1 + \gamma _{k,u,m} x_{k,u,m}\right)^2} - \sum\limits_{a \ne k }^{K_u} {\sum\limits_{n = 1}^{N_{a,u}} { \frac{\beta^{[t]}_{u}w_u W\left[\bOmega_{a,u,u}\right]_{n,m}^2} { \left(\widetilde \gamma _{a,u,n} + {\sigma}^2 +x_{k,u,m}\left[\bOmega_{a,u,u}\right]_{n,m}+\underset{\in \cT_{k, m, a,u}}{\underset{\left(a^{\prime},q,m^{\prime}\right) }{\sum}}x_{a^{\prime},q,m^{\prime}} \left[\bOmega_{a,u,u}\right]_{n,m^{\prime}}\right)^2 }}} \ntb
	&+ \sum\limits_{j \ne u }^{U} {\sum\limits_{a = 1 }^{K_j} {\sum\limits_{n = 1}^{N_{a,j}} { \frac{ \beta^{[t]}_j w_j  W \left[\bOmega_{a,j,u}\right]^2_{n,m}} {\left(\widetilde \gamma _{a,j,n} + {\sigma}^2 + x_{k,u,m}\left[\bOmega_{a,j,u}\right]_{n,m} + \underset{\in \cT_{k, m, a,j}}{\underset{{\left(a^{\prime},q,m^{\prime}\right) }}{\sum} }x_{a^{\prime},q,m^{\prime}} \left[\bOmega_{a,j,u}\right]_{n,m^{\prime}}\right)^2 } } }}
	\end{align}
	\hrule
\end{figure*}

\begin{algorithm}[ht]
	\caption{Centralized Water-Filling Power Allocation Algorithm}
	\label{algCentral}
	\setstretch{1.2}
	\begin{algorithmic}[1]
		\State Initialization: $\bX_{k,u}^0 = \bLambda_{k,u}^{(\ell),[t]}, \; (k,u) \in \cK$, and iteration index $q = 0$. $x_{k,u,m}^q$ stands for the $m$th diagonal entry of $\bX_{k,u}^q$. Initialize $v = 0$ and $\beta_{u}^v = 1/K_u, \mu_{u}^v = 0, u=1,\dots,U$. Set thresholds $\epsilon_3$ and $\epsilon_4$.
		\Repeat
		\For {$u= 1$ to $U$}
		\For {$k= 1$ to $K_u$}
		\For {$m=1$ to $M$}
		\State Set $q' = q$.
		\Repeat
		\State
		Calculate $\varrho_{k,u,m}^{(\ell),[t]}\left(x_{k,u,m}^{q'}\right)$ and  $\varrho_{k,u,m}^{\prime(\ell),[t]}\left(x_{k,u,m}^{q'}\right)$ using \eqref{varrho1} and \eqref{varrho2}, respectively.
		\State
		Update $x^{q'+1}_{k,u,m} = x^{q'}_{k,u,m} - \frac{\varrho_{k,u,m}^{(\ell),[t]}\left(x_{k,u,m}^{q'}\right)}{\varrho_{k,u,m}^{\prime(\ell),[t]}\left(x_{k,u,m}^{q'}\right)}$.
		\State Set $q' = q' + 1$.
		\Until $\left|x^{q'}_{k,u,m} - x^{q'-1}_{k,u,m}\right|\leq \epsilon_3$.
		\EndFor
		\EndFor
		\EndFor
		\State Set $\bar{x}_{k,u,m} = \left[x^{q'}_{k,u,m}\right]^+$ and calculate $p_{\mathrm{tot},u} = \sum\limits_{k=1}^{K_u}\sum\limits_{m=1}^{M}\bar{x}_{k,u,m}$, $u \in \cU$.
		\State Update $\bbeta^{v+1}$ and $\bmu^{v+1}$ by the subgradient method.
		\State Set $v = v +1$.
		\Until $\left\|  \bbeta^v - \bbeta^{v-1}\right\| < \epsilon_4$ and $\left\|  \bmu^v - \bmu^{v-1}\right\| \leq \epsilon_4$.
	\end{algorithmic}
\end{algorithm}

\subsection{Distributed Power Allocation}
In this subsection, considering the backhaul burden caused by the CSI sharing among the BSs in the centralized approach in some cases, we further investigate the distributed approach to find out the power allocation matrices for $\cP_4^{(\ell),[t]}$ in \eqref{eq:problemDinkel}.
For notation convenience, denote the solution of the $(t-1)$th subproblem, i.e., $\cP_4^{(\ell),[t-1]}$, as $\bLambda^{(\ell),[t]}$, and the set of power allocation matrices without cell $u$ after the $(t-1)$th iteration of Dinkelbach's transform as $\bLambda^{(\ell),[t]}_{-u} \triangleq \left\{ {\Lda}^{(\ell),[t]}_{p,q}\: \big| \:\forall (p,q)\in \cK, q\neq u \right\rbrace $.
In the distributed approach, each BS-$u$ optimizes its own power allocation matrices using local CSI, and only a limited number of parameters are shared among cells. Specifically, at the $t$th iteration, each cell optimizes its own power allocation matrices utilizing the set of matrices $\bLambda^{(\ell),[t]}_{-u}$.

Note that the objective function of $\cP_4^{(\ell),[t]}$ is denoted by the minimum of a set of functions with respect to $\bLambda$. When the power allocation matrices of other cells are set to be constant, i.e., represented by $\bLambda^{(\ell),[t]}_{-u}$, the corresponding set of functions becomes
\begin{align}\label{eq:Fu}
F_{u}&\left(\left\lbrace \bLambda_{k,u}\right\rbrace_{k=1}^{K_u} , \bLambda^{(\ell),[t]}_{-u}\right) \ntb
\triangleq  w_u&\overline{R}^{(\ell)}_u\left(\left\lbrace \bLambda_{k,u}\right\rbrace_{k=1}^{K_u}, \bLambda^{(\ell),[t]}_{-u}\right)- \eta^{(\ell),[t]}  P_u\left(\left\lbrace \bLambda_{k,u}\right\rbrace_{k=1}^{K_u} \right) \ntb
= w_u&\sum_{k=1}^{K_u} \left( \log \det \left( \mathbf{I}_{M}+\boldsymbol{\Gamma }_{k,u}{{\bLambda}_{k,u}} \right) - \tr { \mathbf{I}_{N_{k,u}}-{{\boldsymbol{\widetilde{\Phi }}}^{-1}_{k,u}} }\right) \ntb
+& w_u\sum_{k=1}^{K_u} \log \det \left( \boldsymbol{\widetilde{\Gamma }}_{k,u}+\kt\left(\left\lbrace \bLambda_{k,u}\right\rbrace_{k=1}^{K_u}, \Lda^{(\ell),[t]}_{-u} \right) \right) \ntb
-&\eta^{(\ell),[t]} \left(\xi_u \sum\nolimits_{k = 1}^{K_u}{ \tr { \bLambda_{k,u} } } + M\Pc + \Ps \right)\ntb
-& w_u\sum_{k=1}^{K_u}\triangle B_{k,u}^{(\ell)} \left(\left\lbrace \bLambda_{k,u}\right\rbrace_{k=1}^{K_u} , \bLambda^{(\ell),[t]}_{-u}\right), \quad \forall u\in U,
\end{align}
where $F_{u}\left(\left\lbrace \bLambda_{k,u}\right\rbrace_{k=1}^{K_u} , \bLambda^{(\ell),[t]}_{-u}\right)$ is defined for notation convenience.
Note that $F_{u}\left(\left\lbrace \bLambda_{k,u}\right\rbrace_{k=1}^{K_u} , \bLambda^{(\ell),[t]}_{-u}\right)$ is a concave function over each $\bLambda_{k,u}, \; k=1,\dots,K_u$ for given $\bLambda^{(\ell),[t]}_{-u}$.
In addition, denote the optimal solution of problem $\cP_4^{(\ell),[t]}$ in \eqref{eq:problemDinkel} as $\bLambda^{(\ell),[t+1]}$, we can show that the objective functions of problem $\cP_4^{(\ell),[t]}$ tend to be identical \cite{Du2014distriCOMP}, i.e.,
\begin{align}\label{eq:balance}
w_1 W&\overline{R}^{(\ell)}_1\left(\bLambda^{(\ell),[t+1]} \right)- \eta^{(\ell),[t]}  P_1\left(\bLambda^{(\ell),[t+1]}\right) \ntb
= &w_u W\overline{R}^{(\ell)}_u\left(\bLambda^{(\ell),[t+1]} \right)- \eta^{(\ell),[t]}  P_u\left(\bLambda^{(\ell),[t+1]}\right), \ntb
&\qquad \qquad\qquad \qquad u = 2,3,...,U.
\end{align}

Based on this property, we can develop a distributed algorithm where each BS designs its own power allocation matrices individually while sharing limited inter-cell information. For cell $u$, the power allocation problem is cast as
\begin{subequations}\label{eq:problemdistiru}
\begin{align}
\label{eq:distirobj}
\mathcal{P}_{6,u}^{(\ell),[t]}:\; \underset{\left\lbrace \bLambda_{k,u}\right\rbrace_{k=1}^{K_u}} {\max} \quad &  F_{u}\left(\left\lbrace \bLambda_{k,u}\right\rbrace_{k=1}^{K_u} , \bLambda^{(\ell),[t]}_{-u}\right) \\ \label{eq:distirconstraint1}
{\mathrm{s.t.}}\quad
& \sum\limits_{k=1}^{K_u}{\tr{ \bLambda_{k,u} }} \le P_{\max,u}\; \\  \label{eq:distirconstraint2}
& \bLambda_{k,u} \succeq \bzero,\; \bLambda_{k,u} \;\mathrm{diagonal},\; \forall k\in \cK_u,
\end{align}
\end{subequations}
where set $\cK_u \triangleq \left\{1,\dots,K_u\right\}$ denotes all the UTs in cell $u$. We present the following proposition for the solution to problem $\cP_{6,u}^{(\ell),[t]}$.

\begin{prop}\label{propdistri}
Denote the optimal power allocation matrices of problem $\cP_{6,u}^{(\ell),\left[t\right]}$ in \eqref{eq:problemdistiru} as $\bLambda^{(\ell),[t+1]}_{k,u},\;\forall (k,u)\in\cK$. Then, the $m$th diagonal element of $\bLambda_{k,u}^{(\ell),[t+1]}$, i.e., $\lambda_{k,u,m}^{(\ell),[t+1]}$, satisfies \eqref{eq:distrilambda}, shown at the top of next page,
\begin{figure*}
	\begin{equation}\label{eq:distrilambda}
	\left\{ \begin{array}{l}
	\frac{ w_u\gamma_{k,u,m}}{{1 + \gamma _{k,u,m}\lambda^{(\ell),[t+1]}_{k,u,m}}} + \sum\limits_{a \ne k }^{K_u} {\sum\limits_{n = 1}^{N_{a,u}} { \frac{w_u\left[\bOmega_{a,u,u}\right]_{n,m}} { \left[ \boldsymbol{\widetilde{\Gamma }}_{a,u}+\widetilde{\mathbf{K}}_{a,u}\left(\left\lbrace \bLambda_{k,u}^{(\ell),[t+1]}\right\rbrace_{k=1}^{K_u}, \Lda^{(\ell),[t]}_{-u} \right) \right]_{n,n} }} }  = w_u\sum\nolimits_{a \ne k}^{K_u}{d^{(\ell)}_{a,u,u,m}} + \eta^{(\ell),[t]}\xi_u + \theta_u^{[t+1]}, \\
	\quad\quad\quad\quad\quad\quad\quad\quad\quad\quad\quad\quad\quad\quad\quad\quad\quad\quad\quad\quad\quad\quad\quad\quad\quad\quad\quad\quad\quad\quad\quad\quad\quad\quad\quad\quad\quad \theta_u^{[t+1]} < \nu_{k,m}^{(\ell),[t+1]} \\
	\lambda^{(\ell),[t+1]}_{k,u,m} = 0,\quad\quad\quad\quad\quad\quad\quad\quad\quad\quad\quad\quad\quad\quad\quad\quad\quad\quad\quad\quad\quad\quad\quad\quad\quad\quad\quad\quad\quad\;\quad\quad\theta_u^{[t+1]} \geq \nu_{k,m}^{(\ell),[t+1]}
	\end{array} \right.
	\end{equation}
\end{figure*}
where $\lambda^{(\ell),[t+1]}_{k,u,m}$ is the $m$th diagonal elements of $\bLambda_{k,u}^{(\ell),[t+1]}$.
The Lagrange multiplier $\theta_u^{[t+1]}$ is chosen to satisfy
\begin{equation}
	\theta_u^{[t+1]} \left\{\sum_{k=1}^{K_u}{\tr{ \bLambda_{k,u}^{(\ell),[t+1]} }} - P_{\max,u}\right\} = 0, \; \theta_u^{[t+1]} \geq 0,
\end{equation}
and $\nu_{k,m}^{(\ell),[t+1]}$ is the auxiliary variable written as
\begin{align}
&\nu_{k,m}^{(\ell),[t+1]} = w_u\gamma_{k,u,m} -  w_u\sum\limits_{a \ne k}^{K_u}{d^{(\ell)}_{a,u,u,m}} - \eta^{(\ell),[t]}\xi_u  \ntb
& + \sum\limits_{a \ne k }^{K_u} {\sum\limits_{n = 1}^{N_{a,u}} { \frac{w_u\left[\bOmega_{a,u,u}\right]_{n,m}} { \widetilde \gamma _{a,u,n} + \sigma^2 +C_{k,m,a,u}\left(\bLambda^{(\ell),[t]},\bLambda^{(\ell),[t+1]}\right) }} },
\end{align}
where the function $C_{k,m,a,u}\left(\bLambda^{(\ell),[t]},\bLambda^{(\ell),[t+1]}\right)$ and $\mathcal{S}_{k, m, a}$ are defined by \eqref{eq:Cset} and \eqref{eq:Csets} shown at the top of next page, respectively.
\begin{figure*}
	\begin{align}
	C_{k,m,a,u}\left(\bLambda^{(\ell),[t]},\bLambda^{(\ell),[t+1]}\right) = \sum_{q\neq u}^{U}\sum_{p=1}^{K_q} {\tr { \bLambda_{ p,q }^{(\ell),[t]} \widehat \bR_{a,u,q,n}}} + \sum\nolimits_{\left(a^{\prime}, m^{\prime}\right) \in \mathcal{S}_{k, m, a}} \lambda_{a^{\prime},m^{\prime},u}^{(\ell),[t+1]} \left[\bOmega_{a,u,u}\right]_{n,m^{\prime}}\label{eq:Cset}\\
	\mathcal{S}_{k, m, a}=\left\{\left(a^{\prime}, m^{\prime}\right) | a^{\prime} \neq a,\left(a^{\prime}, m^{\prime}\right) \neq(k, m), a^{\prime} \in\{1, \ldots, K_u\}, m^{\prime} \in\{1, \ldots, M\}\right\}\label{eq:Csets}
	\end{align}
\end{figure*}
\end{prop}
\begin{proof}
	See \appref{C}.
\end{proof}

As we can see from \emph{\propref{propdistri}}, the solutions of \eqref{eq:distrilambda} resemble a classical water-filling solutions. Similar to the centralized approach, we apply Newton's method to find the approximate roots of Eq. \eqref{eq:distrilambda}. The auxiliary functions $\rho_{k,m}^{(\ell),[t]}\left(x_{k,m}\right)$ and $\rho_{k,m}^{\prime(\ell),[t]}\left(x_{k,m}\right)$ for Newton's method are defined by \eqref{rho1} and \eqref{rho2} shown at the top of next page, respectively.
\begin{figure*}
\begin{align}\label{rho1}
\rho_{k,m}^{(\ell),[t]}\left(x_{k,m}\right) &= \frac{ w_u\gamma _{k,u,m}}{{1 + \gamma _{k,u,m} x_{k,m}}} -  w_u\sum\limits_{a \ne k}^{K_u}{d^{(\ell)}_{a,u,u,m}} - \eta^{(\ell),[t]}\xi_u - \theta_u^{[t]} \ntb
& + \sum\limits_{a \ne k }^{K_u} {\sum\limits_{n = 1}^{N_{a,u}} { \frac{w_u\left[\bOmega_{a,u,u}\right]_{n,m}} {\widetilde \gamma _{a,u,n} + \sigma^2 + x_{k,m}\left[\bOmega_{a,u,u}\right]_{n,m}+C_{k,m,a,u}\left(\bLambda^{(\ell),[t]},\bX^{s}\right) }}}
\end{align}
\begin{align}\label{rho2}
\rho_{k,m}^{\prime(\ell),[t]}\left(x_{k,m}\right) = -\frac{ w_u\left(\gamma _{k,u,m}\right)^2}{\left(1 + \gamma _{k,u,m} x_{k,m}\right)^2} - \sum\limits_{a \ne k }^{K_u} {\sum\limits_{n = 1}^{N_{a,u}} { \frac{w_u\left[\bOmega_{a,u,u}\right]_{n,m}^2} { \left(\widetilde \gamma_{a,u,n} + \sigma^2 + x_{k,m}\left[\bOmega_{a,u,u}\right]_{n,m}+C_{k,m,a,u}\left(\bLambda^{(\ell),[t]},\bX^{s}\right)\right)^2 }}}
\end{align}
\hrule
\end{figure*}
respectively.
Moreover, the Lagrange multiplier $\theta_u^{[t+1]}$ is updated by the subgradient method.

From \eqref{rho1} and \eqref{eq:Cset}, we can see that for calculating $\rho_{k,m}^{(\ell),[t]}\left(x_{k,m}\right)$, calculations of variables $\eta^{(\ell),[t]}=\underset{u}{\min} \left\{ \eta^{(\ell),[t]}_u \right\}$ and $\sum_{q\neq u}^{U}\sum_{p=1}^{K_q} {\tr { \bLambda_{ p,q }^{(\ell),[t]} \widehat \bR_{a,u,q,n}}}$ require information exchange across BSs, while $\gamma_{k,u,m}$, $w_u$, $d^{(\ell)}_{a,u,u,m}$, $\xi_u$, $\theta_{u}^{[t]}$ and $\bOmega_{a,u,u}$ are all local variables. In addition, the required information exchange for calculating $\rho_{k,m}^{\prime(\ell),[t]}\left(x_{k,m}\right)$ is the same as that for $\rho_{k,m}^{(\ell),[t]}\left(x_{k,m}\right)$.
Then, we write $\sum_{q\neq u}^{U}\sum_{p=1}^{K_q} {\tr { \bLambda_{ p,q }^{(\ell),[t]} \widehat \bR_{a,u,q,n}}}=\sum_{q\neq u}^{U}c_{a,n,q}$
where
	\begin{align}
	c_{a,n,q} = \sum_{p=1}^{K_q}& {\tr { \bLambda_{ p,q }^{(\ell),[t]} \widehat \bR_{a,u,q,n}}}, \ntb
	 &a = 1\dots K_u, a\neq k; n = 1\dots N_{a,u}.
	\end{align}
Note that auxiliary variable $c_{a,n,q}$ is calculated at cell $q\neq u$ and then broadcast to BS-$u$.

Based on the above discussions, we present the distributed iterative water-filling algorithm in \alref{algWaterfilling}. In the following, we quantify the exchanged information of the proposed distributed algorithm. The backhaul signaling mainly happens when calculating the Newton's method auxiliary functions, which corresponds to Step 8 of \alref{algWaterfilling}. In each iteration of the distributed algorithm, the required information to be sent from each BS-$u$ for exchange includes $1 + (U-1) (K_u-1) N_{a,u}$ real-valued scalar variables. So the total number of exchanged variables for all the BSs is $\sum_{u=1}^{U} \left(1 + (U-1) (K_u-1) N_{a,u}\right)$ in each iteration. Note that the proposed distributed algorithm converges within few iterations.
In contrast, for the centralized algorithm, all the statistical CSI, i.e., $\bOmega_{k,u,v} \in \mathbb{R}^{N_{k,u}\times M}$, is required at the central unit for optimization, and then the power allocation results are sent back to all BSs. Then, the required information for exchange includes $M N_{k,u} U \sum_{u=1}^{U}K_u + M \sum_{u=1}^{U}K_u$ real-valued scalar variables. Note that the amount of backhaul signaling of the centralized algorithm scales with the number of antennas $M$, while that of the distributed algorithm does not rely on it. Thus, the proposed distributed algorithm can significantly reduce the backhaul overhead when compared with the centralized one, especially in massive MIMO systems where $M$ is large.

\begin{algorithm}[tbp]
	\caption{Distributed Water-Filling Power Allocation Algorithm}
	\label{algWaterfilling}
	\setstretch{1.2}
	\begin{algorithmic}[1]
		\State Initialization: $\bX_{k}^0 = \bLambda_{k,u}^{(\ell),[t]}, \; k = 1,\dots,K_u$, and iteration index $s = 0$. $x_{k,m}^s$ stands for the $m$th diagonal entry of $\bX_{k}^s$. Initialize $w = 0$ and $\theta_{u}^w = 1/K_u$. Set thresholds $\epsilon_5$ and $\epsilon_6$.
		\Repeat
		\For {$k= 1$ to $K_u$}
		\For {$m=1$ to $M$}
		\State Set $s' = s$.
		\Repeat
		\State BS-$u$ receives $c_{a,n,q}, \; q \neq u; a = 1\dots K_u, a\neq k; n = 1\dots N_{a,u}$.
		\State
		Calculate $\rho_{k,m}^{(\ell),[t]}\left(x_{k,m}^{s'}\right)$ and  $\rho_{k,m}^{\prime(\ell),[t]}\left(x_{k,m}^{s'}\right)$ with \eqref{rho1} and \eqref{rho2}, respectively.
		\State
		Update $x^{s'+1}_{k,m} = x^{s'}_{k,m} - \frac{\rho_{k,m}^{(\ell),[t]}\left(x_{k,m}^{s'}\right)}{\rho_{k,m}^{\prime(\ell),[t]}\left(x_{k,m}^{s'}\right)}$.
		\State $s' = s' + 1$.
		\Until $\left|x^{s'}_{k,m} - x^{s'-1}_{k,m}\right|\leq \epsilon_5$.
		\EndFor
		\EndFor
		\State Set $\bar{x}_{k,m} = \left[x^{s'}_{k,m}\right]^+$ and calculate $p_{\mathrm{tot},u} = \sum_{k,m}\bar{x}_{k,m}$, $k=1,\dots,K_u;m=1,\dots,M$.
		\State Update $\theta_u^{w+1}$ by the subgradient method.
		\Until $\left|\theta^{w}_u - \theta^{w-1}_u\right|\leq \epsilon_6$.
	\end{algorithmic}
\end{algorithm}

\subsection{Complexity Analysis}
We analyze the complexity of the proposed algorithms in this subsection. For \alref{algMMDinkel}, the main complexity of each iteration lies in the complexity of tackling problem $\cP_4^{(\ell),[t]}$ in \eqref{eq:problemDinkel}, which is solved by the centralized \alref{algCentral} or by the distributed \alref{algWaterfilling}. In the following, we analyze the complexity for the proposed centralized and distributed approaches, respectively.

For the centralized approach, the major complexity of \alref{algCentral} lies in two aspects, i.e., solving Eq. \eqref{eq:newtoncentral} via Newton's method and updating the Lagrange multipliers through the subgradient method. Specifically, with the precision set to be $d$ digits \cite{Boyd04Convex}, the required number of iterations for Newton's method is $\log d$ \cite{Cormen09Algorithm}.
Then, the computational complexity of \alref{algCentral} is $\cO(L_{\mathrm{S}} U K M \left(\log d +1\right))$ \cite{wu18beam}, where $L_{\mathrm{S}}$ is the number of iterations in the subgradient method, and $K = \sum_{u=1}^{U}K_u$.
Therefore, the computational complexity of the centralized method is approximately  $\cO(L_{\mathrm{M}} L_{\mathrm{D}} L_{\mathrm{S}} U K M \left(\log d +1\right))$, where $L_{\mathrm{M}}$ and $L_{\mathrm{D}}$ are the required numbers of iterations for the MM method, Dinkelbach's transform, respectively.

For the distributed approach where the problem $\cP_4^{(\ell),[t]}$ is solved by \alref{algWaterfilling}, the complexity analysis of \alref{algWaterfilling} is similar to that of \alref{algCentral}, which is $\cO(L_{\mathrm{S}} K_u M \left(\log d +1\right))$. Then we can formulate the complexity of the distributed algorithm as $\cO(L_{\mathrm{M}} L_{\mathrm{D}} L_{\mathrm{S}} K M \left(\log d +1\right))$, which is reduced compared with the centralized algorithm, especially when $U$ is large. Moreover, the distributed approach can perform power allocation with limited intercell coordination at an acceptable sacrifice on the EE performance compared with the centralized one, which will be shown in \secref{sec:numerical} by numerical results.
Note that if problem $\cP_4^{(\ell),[t]}$ in \eqref{eq:problemDinkel} of \alref{algMMDinkel} is solved via the interior point method, the complexity can reach $\cO(L_{\mathrm{M}} L_{\mathrm{D}} K^3 M^3)$ \cite{Boyd04Convex}. So our proposed algorithm can significantly reduce the computational complexity.

\section{Numerical Results}\label{sec:numerical}
We provide numerical analysis to illustrate the performance of the proposed max-min EE iterative algorithms in this section. The WINNER \Rmnum{2} channel model is employed to conduct the simulations \cite{Winner2}. Note that the suburban scenario with non-line-of-sight (NLOS) propagation is considered.
In our simulations, the number of cells is $U=3$, each cell contains one BS with $M=128$ antennas and $K_u=4$ UTs with $N_{k,u}=4$ antennas. Both the BS and the UTs are equipped with ULAs where the spacing between antennas is half-wavelength. We set the noise power $\sigma^2$ as $-105$ dBm, the weighting factor as $w_u=1, \forall u$,  $\Pc$ and $\Ps$ are set as $30$ dBm and $40$ dBm, respectively. The amplifier inefficiency factor is set as $\xi_u=5,\forall u$. The large scale fading factor is given as $10\log_{10}\left(\theta_{k,u,v}\right) = -38\log_{10}\left(d_{k,u,v}-34.5+c_{k,u,v}\right)$, where $d_{k,u,v}$ represents the distance between the BS-$v$ and UT-$(k,u)$ and $c_{k,u,v}$ stands for the log-normal shadow fading with zero mean and $8$ dB deviation \cite{He2013coorbeam}. The power budgets are equal for all cells, i.e., $P_{\max,u}=P_{\max}$, $\forall u$.

\begin{figure}
	\centering
	\includegraphics[width=0.45\textwidth]{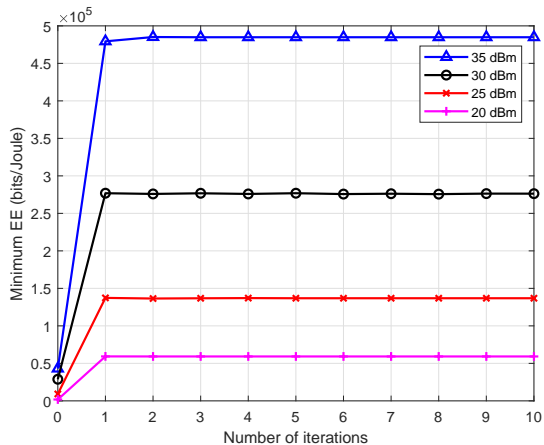}
	\caption{Convergence behavior of \textbf{Algorithm 1} versus the numbers of iterations for different values of power budgets $P_{\mathrm{max}}$.}
	\label{conver}
\end{figure}

Firstly, we evaluate the convergence performance of \alref{algMMDinkel} in Fig. \ref{conver}. The results indicate that the proposed \alref{algMMDinkel} converges quickly after about two or three iterations under different power budgets.

\begin{figure}
	\centering
	\includegraphics[width=0.45\textwidth]{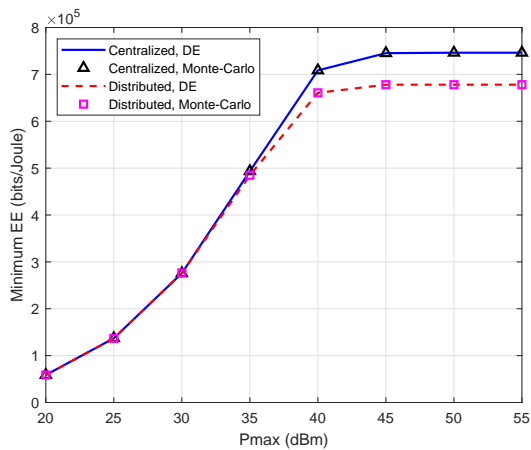}
	\caption{The minimum EE performance of the distributed and centralized algorithms.}
	\label{discen}
\end{figure}

Next, the EE performance of the proposed distributed and centralized power allocation algorithms are compared in \figref{discen}. We can observe that as the power budget gets higher, both the minimum EE of the distributed and centralized algorithms become higher, and eventually reach a stationary point. In addition, in the low power budget regime, the noise dominates the interference. Then, the EE performance of centralized and distributed approaches are almost identical. While in the higher power budget regime, the centralized algorithm outperforms the distributed one in terms of the minimum EE performance as the noise becomes negligible compared to the interference. We can also find that the DE results match Monte-Carlo results accurately.

\begin{figure}
	
	\centering
	\subfigure [Minimum EE]{
			\centering
			\includegraphics[width=0.45\textwidth]{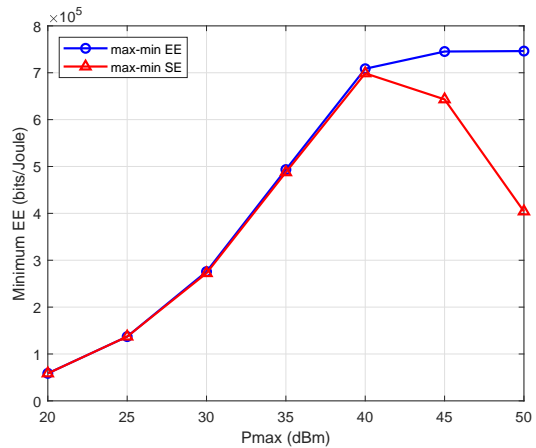}
			\label{fig_minEE}
	}
	\subfigure [Minimum SE]{
			\centering
			\includegraphics[width=0.45\textwidth]{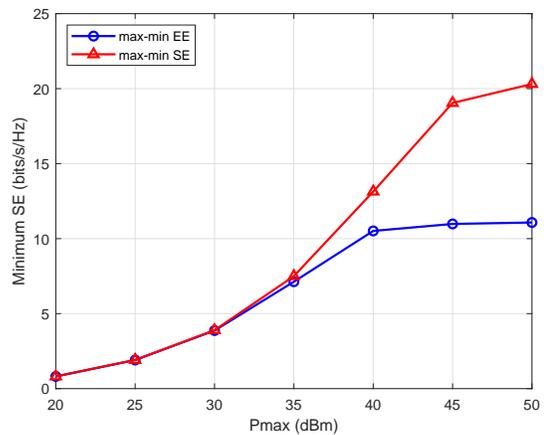}
			\label{fig_minSE}
	}
	\caption{Comparison between the proposed max-min EE and max-min SE approaches.}\label{fig_EESEcompare}
\end{figure}

In \figref{fig_EESEcompare}, we compare our proposed max-min EE approach to the max-min SE approach, which is carried out by letting $\xi_u = 0, \forall u$ in the proposed max-min EE method. In particular, we show the minimum EE and minimum SE performance versus the transmit power constraint of the two methods in \figref{fig_minEE} and \figref{fig_minSE}, respectively.
We can observe that the max-min EE and max-min SE approaches achieve almost identical performance when the power budget is low, which indicates that the minimum SE optimal is nearly minimum EE optimal when transmitting with a full power budget.
Nevertheless, in the higher power budget regime, the EE value of the max-min EE method tends to saturate, while that of the max-min SE approach starts to decline rapidly, as depicted in \figref{fig_minEE}. Moreover, in \figref{fig_minSE}, the minimum SE value of the max-min SE method continues to grow, and that of max-min EE method tends to saturate.
This is due to the existence of a threshold for the transmit power to maximize the minimum EE. When the transmission power budget exceeds that threshold, the system EE will not increase. On the contrary, the max-min SE method tends to optimize the minimum SE regardless of the power used, which might lead to a decline of the system EE.

\begin{figure}
	\centering
	\includegraphics[width=0.45\textwidth]{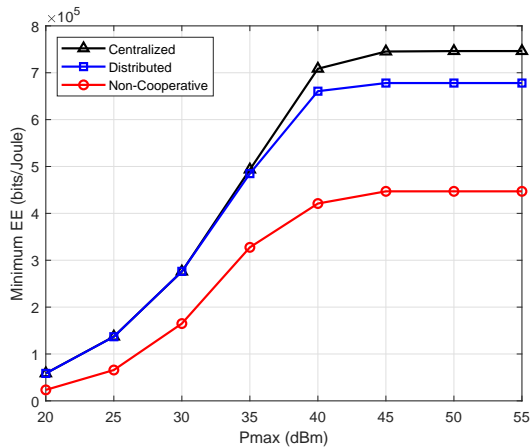}
	\caption{Comparison between the proposed centralized and distributed approaches and the non-cooperative baseline.}
	\label{noncoorp}
\end{figure}

\figref{noncoorp} compares the minimum EE performance of our proposed centralized and distributed approaches with the conventional non-cooperative baseline, where each cell conducts its own precoding design under the criterion of maximizing the minimum EE of the UTs in its cell, regardless of the interference signal from other cells.
We can observe that our proposed algorithms achieve better performance than the non-cooperative baseline. This is because during the cell-cooperation, precoding procedures are jointly processed among cells on the basis of information sharing and then the ICI is greatly mitigated.

\begin{figure}
	\centering
	\subfigure[$w_1 = w_2 = w_3$]{
			\centering
			\includegraphics[width=0.14\textwidth]{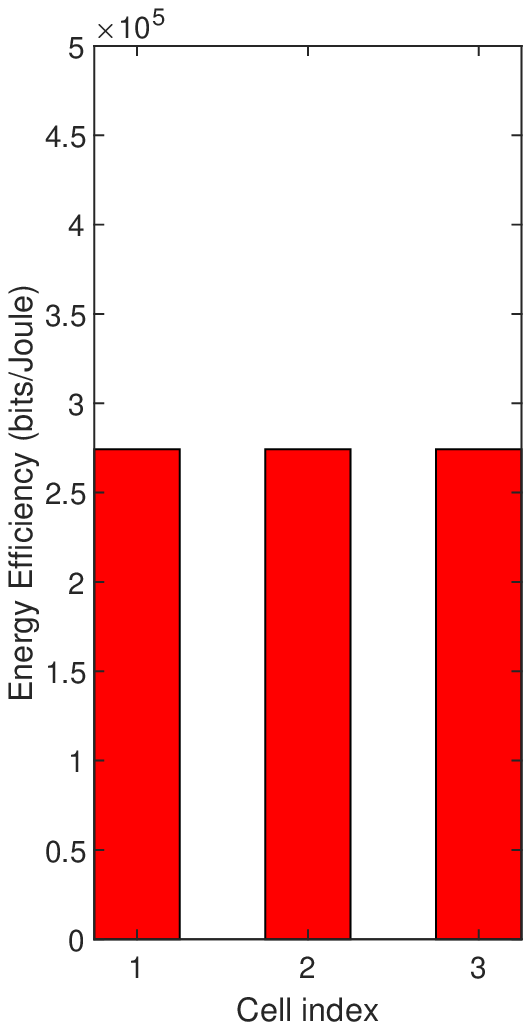}
			\label{weightfactor1}
	}
	\subfigure[$w_1 = 0.5w_2 = 2w_3$]{
		\centering
		\includegraphics[width=0.14\textwidth]{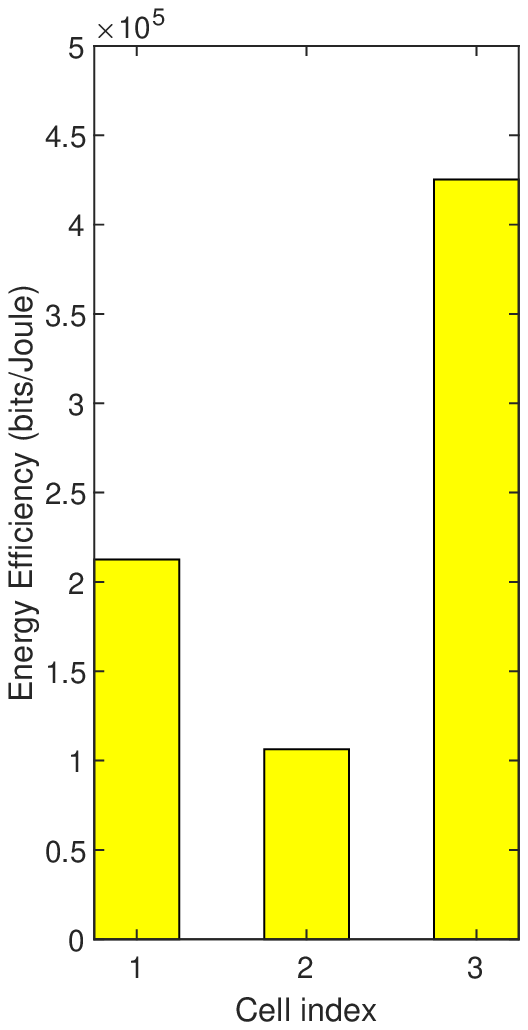}
		\label{weightfactor3}
    }
	\subfigure[$0.25w_1 = w_2 = 0.5w_3$]{
		\centering
		\includegraphics[width=0.14\textwidth]{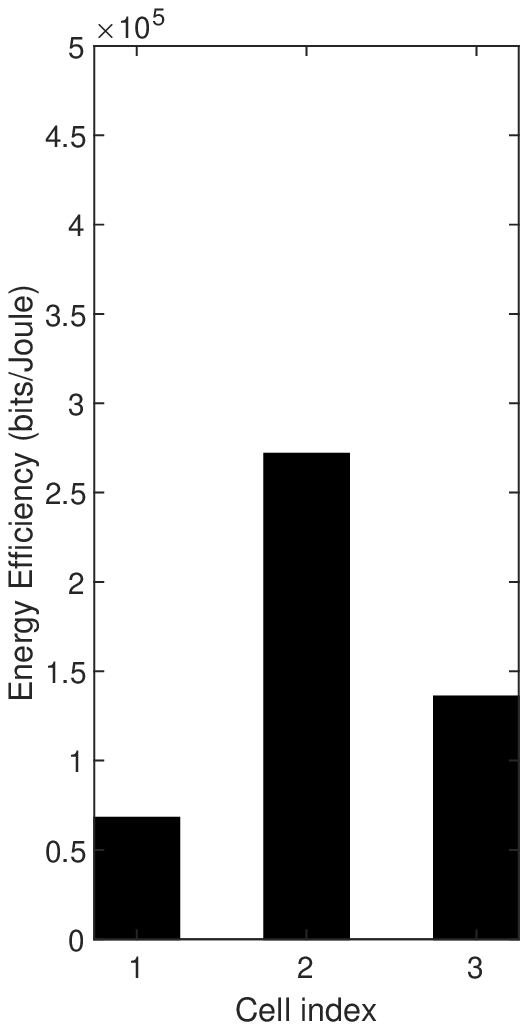}
		\label{weightfactor4}
    }
	\caption{The impacts of weighting factor on EE distribution under the power constraint $P_{\mathrm{max},u} = 30$ dBm, $\forall u\in \cU$.}
	\label{weightfactor}
\end{figure}

The impact of the weighting factors on the distribution of the EE performance is shown in \figref{weightfactor}.
For the equal weighting factor case where $w_u = 1, \forall u\in \cU$, as shown in \figref{weightfactor1}, the EE is almost identical among the cells. We further evaluate the cases when the weighting factors are unequal. In particular, we consider the cases with $w_1 = 0.5w_2 = 2w_3$ and $0.25w_1 = w_2 = 0.5w_3$ in \figref{weightfactor3} and \figref{weightfactor4}, respectively. We can observe that the EE performance differs from each other according to their weights, which offers us a way to control the EE distribution between the cells when specific EE demands are required for some cells.

\begin{figure}
	\centering
	\includegraphics[width=0.45\textwidth]{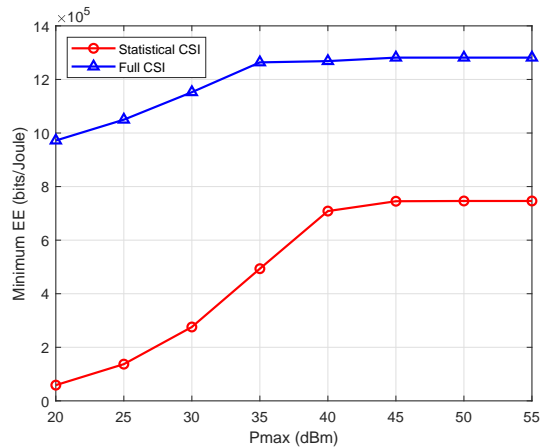}
	\caption{The comparison between the statistical CSI and full CSI approaches on minimum EE performance.}
	\label{insCompare}
\end{figure}

Finally, the EE performance of the proposed power allocation approach and that of the full CSI approach which assumes instantaneous CSI is available at each BS are compared in \figref{insCompare}. Note that although the performance of the instantaneous CSI approach is better than our statistical CSI based one, more signalling overhead and higher complexity is incurred.

\section{Conclusion}\label{sec:conclusion}
We studied the precoding design in multi-cell multi-user massive MIMO downlink systems utilizing statistical CSI. Our objective was to maximize the minimum weighted EE by designing the transmit covariance matrices under the per BS power constraint. We first exploited the massive MIMO channel properties to derive the closed-form optimal transmit directions at each BS, thus simplifying the precoding design problem into a power allocation one.
We adopted a asymptotically accurate surrogate expression of the ergodic user rate, i.e., its DE, which significantly reduced the optimization complexity. We further applied the MM technique and Dinkelbach's transform to deal with this max-min fractional optimization problem.
Furthermore, we proposed two methods to address the obtained subproblems, i.e., the centralized and the distributed ones. For each method, we proposed generalized water-filling schemes to obtain the solution. Numerical results demonstrated the performance of the proposed algorithms.

\begin{appendices}

\section{Proof of Proposition 2}\label{B}
The Lagrangian function of problem $\mathcal{P}_5^{(\ell),[t]}$ is
\begin{align}\label{eq:dislagranorigin}
{\cal L} &\left( z,\bLambda,\bbeta,\bmu \right)  =  z - \sum\limits_{u=1}^{U}{  \mu_u \left\{\sum\limits_{k=1}^{K_u}{\tr{ \bLambda_{k,u} }} - P_{\max,u}\right\} } \ntb
& - \sum\limits_{u=1}^{U}{ \beta_u \left\{  z -  \ w_u \left\{   W \overline{R}^{(\ell)}_u\left(\bLambda \right)- \eta^{(\ell),[t]}  P_u\left(\bLambda\right)  \right\}  \right\} }\ntb
& = \left( 1 - \sum\limits_{u=1}^{U}{\beta_u}  \right)z + \sum\limits_{u=1}^{U} \mu_u \left( { P_{\max,u}} - \sum\limits_{k=1}^{K_u}{\tr{ \bLambda_{k,u} }} \right) \ntb
& \qquad + \sum\limits_{u=1}^{U} {\beta_u}w_u { \left(   W \overline{R}^{(\ell)}_u\left(\bLambda \right)- \eta^{(\ell),[t]}  P_u\left(\bLambda\right) \right) },
\end{align}
where the Lagrange multipliers $\bbeta = \left\{ \beta_1,\ldots,\beta_U \right\},\beta_u \ge 0 \ (\forall u)$ and $\bmu = \left\{ \mu_1,\ldots,\mu_U \right\},\mu_u \ge 0 \ (\forall u)$ are associated with the dual variables corresponding to the constraints in \eqref{eq:problemzconstraint1} and \eqref{eq:problemzconstraint2}, respectively.

To maximize $\cL$ in \eqref{eq:dislagranorigin} with fixed $\bbeta$ and $\bmu$, the following conditions have to be satisfied
\begin{subequations}\label{eq:KKT}
	\begin{align}\label{eq:KKT1}
	\frac{\partial \cL }{\partial z} & = 1 - \sum\limits_{u=1}^{U}{\beta_u} =0, \\\label{eq:KKT2}
	\frac{\partial \cL }{\partial \bLambda_{k,u} } & = \sum\limits_{j=1}^{U} \beta_j w_j \left(  W \frac{\partial \overline{R}^{(\ell)}_j\left(\bLambda \right) }{\partial \bLambda_{k,u} } - \ \eta^{(\ell),[t]}\xi_j \bI_M \right)\ntb
	 & \qquad -\sum\limits_{j=1}^{U} \mu_j \bI_M  = \bzero, \quad \forall (k,u) \in \cK.
	\end{align}
\end{subequations}
It can be observed from \eqref{eq:KKT1} that to obtain the maximal value of $\cL$, the summation of $\beta_u$ has to be equal to one. By plugging \eqref{eq:KKT1} into \eqref{eq:dislagranorigin}, we can find that $z$ will not influence the value of $\cL$, which further leads to the simplification of $\cL\left( z,\bLambda,\bbeta,\bmu \right)$ to $\cL\left( \bLambda,\bbeta,\bmu \right)$.

To derive the gradient of $\overline{R}^{(\ell)}_j\left(\bLambda \right)$ over $\bLambda_{k,u}$, we rewrite the derivative $\frac{\partial \overline{R}^{(\ell)}_j\left(\bLambda \right) }{\partial \bLambda_{k,u} }$ in \eqref{eq:KKT2} as
\begin{align}\label{eq:der}
\frac{\partial \overline{R}^{(\ell)}_j\left(\bLambda \right) }{\partial \bLambda_{k,u} }  = \frac{\partial }{\partial { \bLambda_{k,u} }}\sum\limits_{a = 1}^{K_j} { \Abar_{a,j} \left(\Lda\right)} - \frac{\partial }{\partial { \bLambda_{k,u} }}\sum\limits_{a = 1}^{K_j} { \triangle B_{a,j}^{(\ell)}\left( \Lda \right) },
\end{align}
where the gradient of ${ \Abar_{a,j} \left(\Lda\right)}$ and ${ \triangle B_{a,j}^{(\ell)}\left( \Lda \right) }$ over $\bLambda_{k,u}$ are described by \eqref{eq:der_a} and \eqref{eq:der_b}, respectively, shown at the top of next page,
\begin{figure*}
	\begin{align}\label{eq:der_a}
	\frac{\partial  }{\partial { \bLambda_{k,u} }}\sum\limits_{a = 1}^{K_j} { \Abar_{a,j} \left(\Lda\right)}  & = \left\{ \begin{array}{l}
	\left( \bI_{M} + {\bf{\Gamma }}_{k,u} \bLambda_{k,u} \right)^{ - 1} {\bf{\Gamma }}_{k,u} + \sum\limits_{a \ne k }^{K_u} {\sum\limits_{n = 1}^{N_{a,u}} { \frac{\widehat \bR_{a,u,u,n}} {\widetilde \gamma _{a,u,n} + {\sigma}^2 + \sum\nolimits_{(p,q) \ne (a,u) } {\tr { \bLambda_{ p,q } \widehat \bR_{a,u,q,n}}} }} }, \quad j = u \\
	\sum\limits_{a = 1 }^{K_j} {\sum\limits_{n = 1}^{N_{a,j}} { \frac{\widehat \bR_{a,j,u,n}} {\widetilde \gamma _{a,j,n} + {\sigma}^2 +  \sum\nolimits_{(p,q) \ne (a,j) }{ \tr { \bLambda_{p,q } \widehat \bR_{a,j,q,n}}}} } },\quad\quad\quad\quad\qquad \quad\quad\quad\qquad\qquad \ j \ne u
	\end{array} \right. \\\label{eq:der_b}
	\frac{\partial  }{\partial { \bLambda_{k,u} }}\sum\limits_{a = 1}^{K_j} { \triangle B_{a,j}^{(\ell)} \left(\Lda\right)} & =  \left\{ \begin{array}{l}
	\frac{\partial }{\partial { \bLambda_{k,u} }}\sum\nolimits_{a \ne k}^{K_u} { B_{a,u}\left(\Lda^{(\ell)}\right) }  = \sum\limits_{a \ne k}^{K_u} \bD^{(\ell)}_{a,u,u}, \quad j = u \\
	\frac{\partial }{\partial { \bLambda_{k,u} }} \sum\nolimits_{a = 1}^{K_j} { B_{a,j}\left(\Lda^{(\ell)}\right) } = \sum\limits_{a = 1}^{K_j} \bD^{(\ell)}_{a,j,u}, \quad \  j \ne u
	\end{array} \right.
	\end{align}
\end{figure*}
and $\widetilde \gamma _{a,u,n}$ represents the $n$th diagonal element of $\widetilde{\mathbf{\Gamma } }_{a,u}$.

With the aid of \eqref{eq:der}, we can rewrite the KKT condition \eqref{eq:KKT2} as
\begin{align}\label{eq:reKKT2}
&\frac{\partial \cL }{\partial \bLambda_{k,u} }  = \beta_u w_u  W \left( \bI_{M} + {\bf{\Gamma }}_{k,u} \bLambda_{k,u} \right)^{ - 1}{\bf{\Gamma }}_{k,u} \ntb
& + \sum\limits_{a \ne k }^{K_u} {\sum\limits_{n = 1}^{N_{a,u}} { \frac{\beta_u w_u  W \widehat \bR_{a,u,u,n}} {\widetilde \gamma _{a,u,n} + {\sigma}^2 + \sum\nolimits_{(p,q) \ne (a,u) } {\tr { \bLambda_{ p,q } \widehat \bR_{a,u,q,n}}} }} } \ntb
& + \sum\limits_{j \ne u }^{U} {\sum\limits_{a = 1 }^{K_j} {\sum\limits_{n = 1}^{N_{a,j}} { \frac{ \beta_j w_j  W \widehat \bR_{a,j,u,n}} {\widetilde \gamma _{a,j,n} + {\sigma}^2 +  \sum\nolimits_{(p,q) \ne (a,j) }{ \tr { \bLambda_{p,q } \widehat \bR_{a,j,q,n}}}} } }} \ntb
& - \sum\limits_{j \ne u }^{U} {\beta_j w_j  W \sum\limits_{a = 1}^{K_j} \bD^{(\ell)}_{a,j,u} }- \beta_u w_u  W \sum\limits_{a \ne k}^{K_u} \bD^{(\ell)}_{a,u,u} \ntb
& - \sum\limits_{j=1}^{U}{ \left( \ \beta_j w_j \eta^{(\ell),[t]}\xi_j + \mu_j\right) \bI_M } = \bzero, \quad \forall (k,u) \in \cK.
\end{align}
Note that $\frac{\partial \cL }{\partial \bLambda_{k,u} }$ is a diagonal matrix. Then, the KKT condition in \eqref{eq:reKKT2} can be further reduced to
\begin{align}\label{eq:KKT_app}
&\left[ \frac{\partial \cL }{\partial \bLambda_{k,u} } \right]_{m,m}   = {\frac{ \beta_u w_u  W \gamma _{k,u,m}}{{1 + \gamma _{k,u,m}\lambda _{k,u,m}}}} - \beta_u w_u  W \sum\limits_{a \ne k}^{K_u}{d^{(\ell)}_{a,u,u,m}} \ntb
& + \sum\limits_{a \ne k }^{K_u} {\sum\limits_{n = 1}^{N_{a,u}} { \frac{\beta_u w_u  W \left[\bOmega_{a,u,u}\right]_{n,m}} {\widetilde \gamma _{a,u,n} + {\sigma}^2 + \sum\nolimits_{(p,q) \ne (a,u) } {\tr { \bLambda_{ p,q } \widehat \bR_{a,u,q,n}}} }} }\ntb
& + \sum\limits_{j \ne u }^{U} {\sum\limits_{a = 1 }^{K_j} {\sum\limits_{n = 1}^{N_{a,j}} { \frac{ \beta_j w_j  W \left[\bOmega_{a,j,u}\right]_{n,m}} {\widetilde \gamma _{a,j,n} + {\sigma}^2 +  \sum\nolimits_{(p,q) \ne (a,j) }{ \tr { \bLambda_{p,q } \widehat \bR_{a,j,q,n}}}} } }}\ntb
& - \sum\limits_{j=1}^{U}{ \left( \ \beta_j w_j \eta^{(\ell),[t]}\xi_j + \mu_j\right) } - \sum\limits_{j \ne u }^{U} {\beta_j w_j  W \sum\limits_{a = 1}^{K_j} d^{(\ell)}_{a,j,u,m} }=0, \ntb
&\qquad\qquad\quad\quad \qquad\qquad\quad\quad\qquad\qquad m = 1, \ldots ,M.
\end{align}

Solving the KKT conditions in \eqref{eq:KKT_app}, we can find that the $m$th diagonal element of  $\bLambda_{k,u}^{(\ell),[t+1]}, \forall (k,u)\in\cK$ satisfies \eqref{eq:cenkkt} shown at the top of next page,
\begin{figure*}
	\begin{equation}\label{eq:cenkkt}
	\left\{ \begin{array}{l}
	\frac{ \beta_{u} w_u W\gamma_{k,u,m}}{{1 + \gamma _{k,u,m}\lambda^{(\ell),[t+1]}_{k,u,m}}}
	+ \sum\limits_{a \ne k }^{K_u} {\sum\limits_{n = 1}^{N_{a,u}}  \frac{\beta_u w_u  W \left[\bOmega_{a,u,u}\right]_{n,m}} {\widetilde \gamma _{a,u,n} + \left[ \widetilde{\mathbf{K}}_{a,u} (\bLambda) \right]_{n,n} }  }
	+ \sum\limits_{j \ne u }^{U} {\sum\limits_{a = 1 }^{K_j} {\sum\limits_{n = 1}^{N_{a,j}} { \frac{ \beta_j w_j  W \left[\bOmega_{a,j,u}\right]_{n,m}} {\widetilde \gamma _{a,j,n} + \left[ \widetilde{\mathbf{K}}_{a,j} (\bLambda) \right]_{n,n} } } }} = \eta^{(\ell),[t]}\xi_u + \mu_u^{[t+1]}\\
	\quad \quad \quad\quad \quad \quad\quad \quad + \beta_{u} w_u W \sum\limits_{a \ne k}^{K_u}{d^{(\ell)}_{a,u,u,m}} + \sum\limits_{j \ne u }^{U} {\beta_j w_j  W \sum\limits_{a = 1}^{K_j} d^{(\ell)}_{a,j,u,m} }, \quad \quad\quad
	\mu_u^{[t+1]} < \chi_{k,m}^{(\ell),[t+1]}\\
	\lambda^{(\ell),[t+1]}_{k,u,m} = 0,\quad\quad\quad\quad\quad\quad\quad\quad\quad\quad\quad\quad\quad\quad\quad\quad\quad \quad\quad\quad\quad\quad\quad\quad\quad\quad \mu_u^{[t+1]} \geq \chi_{k,m}^{(\ell),[t+1]}
	\end{array} \right.
	\end{equation}
\end{figure*}
where the auxiliary variables $\chi_{k,m}^{(\ell),[t+1]}$ and $\cT_{k, m, a,u}$ are written as \eqref{eq:cenaux} and \eqref{eq:cenauxtt} shown at the top of next page, respectively.
\begin{figure*}
\begin{align}
	 \chi_{k,m}^{(\ell),[t+1]} &= \gamma_{k,u,m} -  \beta_{u} w_u W \sum\limits_{a \ne k}^{K_u}{d^{(\ell)}_{a,u,u,m}} + \sum\limits_{a \ne k }^{K_u} {\sum\limits_{n = 1}^{N_{a,u}} { \frac{\beta_u w_u  W \left[\bOmega_{a,u,u}\right]_{n,m}} {\widetilde \gamma _{a,u,n} + {\sigma}^2 + \sum\nolimits_{\left(a^{\prime},q,m^{\prime}\right) \in \cT_{k, m, a,u}} \lambda_{a^{\prime},q,m^{\prime}}^{(\ell),[t+1]} \left[\bOmega_{a,u,u}\right]_{n,m^{\prime}} }} } \ntb
	&- \eta^{(\ell),[t]}\xi_u  -\sum\limits_{j \ne u }^{U} {\beta_j w_j  W \sum\limits_{a = 1}^{K_j} d^{(\ell)}_{a,j,u,m} } + \sum\limits_{j \ne u }^{U} {\sum\limits_{a = 1 }^{K_j} {\sum\limits_{n = 1}^{N_{a,j}} { \frac{ \beta_j w_j  W \left[\bOmega_{a,j,u}\right]_{n,m}} {\widetilde \gamma _{a,j,n} + {\sigma}^2 +  \sum\nolimits_{\left(a^{\prime},q,m^{\prime}\right) \in \cT_{k, m, a,j}} \lambda_{a^{\prime},q,m^{\prime}}^{(\ell),[t+1]} \left[\bOmega_{a,j,u}\right]_{n,m^{\prime}}} } }}\label{eq:cenaux}\\
	\cT_{k, m, a,u} &=\left\{\left(a^{\prime},q,m^{\prime}\right) | \left(a^{\prime},q\right) \neq (a,u),\left(a^{\prime}, q,m^{\prime}\right) \neq(k,u, m),\left(a^{\prime},q,m^{\prime}\right) \in\cK, m^{\prime} \in\{1, \ldots, M\}\right\}\label{eq:cenauxtt}
\end{align}
\hrule
\end{figure*}
This concludes the proof.

\section{Proof of Proposition 3}\label{C}
The Lagrangian function of problem $\mathcal{P}_{6,u}^{(\ell),[t]}$ is
\begin{align}\label{eq:dislag1}
{\cal L} &\left( \left\lbrace \bLambda_{k,u}\right\rbrace_{k=1}^{K_u},\theta_u \right)  =  F_{u} \left(\left\lbrace \bLambda_{k,u}\right\rbrace_{k=1}^{K_u} , \bLambda^{(\ell),[t]}_{-u}\right) \ntb
&+\sum\limits_{k=1}^{K_u} \operatorname{tr}\left(\mathbf{A}_{k,u} \mathbf{\Lambda}_{k,u}\right) \theta_u \left\{\sum\limits_{k=1}^{K_u}{\tr{ \bLambda_{k,u} }} - P_{\max,u}\right\},
\end{align}
where the Lagrange multiplier $\left\lbrace \bA_{k,u}\succeq 0 \right\rbrace $ and $\theta_u \geq 0$ are associated with the problem constraints in \eqref{eq:distirconstraint1} and \eqref{eq:distirconstraint2}, respectively.

The KKT conditions of \eqref{eq:problemdistiru} are
\begin{align}
\label{eq:KKT1dis}
&\frac{\partial \cL }{\partial \bLambda_{k,u} } = \frac{\partial F_{u} }{\partial \bLambda_{k,u}}  - \theta_u \bI_M  = \bzero, \quad \forall k \in \cK_u,\\ \label{eq:KKT2dis}
&\theta_u \left\{\sum_{k=1}^{K_u}{\tr{ \bLambda_{k,u} }} - P_{\max,u}\right\} = 0, \quad \theta_u \geq 0, \\ \label{eq:KKT3dis}
&\tr{\mathbf{A}_{k,u} \mathbf{\Lambda}_{k,u}}=0, \quad \mathbf{A}_{k,u} \succeq \mathbf{0}, \quad \Lambda_{k,u} \succeq \mathbf{0}.
\end{align}
Since \eqref{eq:problemdistiru} is a convex optimization problem, we can solve the KKT conditions to find the optimal solution. Calculate the derivative of  $F_{u}\left(\left\lbrace \bLambda_{k,u}\right\rbrace_{k=1}^{K_u} , \bLambda^{(\ell),[t]}_{-u}\right)$ over the transmit power matrices of cell $u$, the first KKT condition in \eqref{eq:KKT1dis} can be rewritten as
\begin{align}
&\frac{\partial \cL }{\partial \bLambda_{k,u} } = \frac{\partial F_{u} }{\partial \bLambda_{k,u}}  - \theta_u \bI_M = w_u\left( \bI_{M} + {\boldsymbol{\Gamma }}_{k,u} \bLambda_{k,u} \right)^{ - 1} {\boldsymbol{\Gamma }}_{k,u}\ntb
&+ \sum\limits_{a \ne k }^{K_u} {\sum\limits_{n = 1}^{N_{a,u}} { \frac{w_u\widehat \bR_{a,u,u,n}} { \left[ \boldsymbol{\widetilde{\Gamma }}_{a,u}+\widetilde{\mathbf{K}}_{a,u}\left(\left\lbrace \bLambda_{a,u}\right\rbrace_{a=1}^{K_u}, \Lda^{(\ell),[t]}_{-u} \right) \right]_{n,n} }} }   \ntb
& - w_u\sum\limits_{a \ne k}^{K_u} \bD^{(\ell)}_{a,u,u} - \left(\eta^{(\ell),[t]}\xi_u+\theta_u\right) \bI_M = \bzero, k \in \cK_u.
\end{align}

Using the KKT conditions in \eqref{eq:KKT1dis}--\eqref{eq:KKT3dis}, the solution of $\mathcal{P}_{6,u}^{(\ell),[t]}$, which is $\left\lbrace \bLambda_{k,u}^{(\ell),[t+1]}\right\rbrace_{k=1}^{K_u}$, satisfies \eqref{eq:diskkt} shown at the top of next page,
\begin{figure*}
	\begin{equation}\label{eq:diskkt}
	\left\{ \begin{array}{l}
	\frac{ w_u\gamma_{k,u,m}}{{1 + \gamma _{k,u,m}\lambda^{(\ell),[t+1]}_{k,u,m}}} + \sum\limits_{a \ne k }^{K_u} {\sum\limits_{n = 1}^{N_{a,u}} { \frac{w_u\left[\bOmega_{a,u,u}\right]_{n,m}} { \left[ \boldsymbol{\widetilde{\Gamma }}_{a,u}+\widetilde{\mathbf{K}}_{a,u}\left(\left\lbrace \bLambda_{k,u}^{(\ell),[t+1]}\right\rbrace_{k=1}^{K_u}, \Lda^{(\ell),[t]}_{-u} \right) \right]_{n,n} }} } \\
	\quad\quad\quad\quad\quad\quad\;\, = w_u\sum\nolimits_{a \ne k}^{K_u}{d^{(\ell)}_{a,u,u,m}} + \eta^{(\ell),[t]}\xi_u + \theta_u^{[t+1]},\quad \quad\quad
	\theta_u^{[t+1]} < \nu_{k,m}^{(\ell),[t+1]} \\
	\lambda^{(\ell),[t+1]}_{k,u,m} = 0,\quad\quad\quad\quad\quad\quad\quad\quad\quad\quad\quad\quad\quad\quad\quad\quad\quad\quad\quad\quad\quad\theta_u^{[t+1]} \geq \nu_{k,m}^{(\ell),[t+1]}
	\end{array} \right.
	\end{equation}
	\hrule
\end{figure*}
where $\theta_u^{[t+1]}$ is chosen to satisfy the KKT condition in \eqref{eq:KKT2dis}, and the auxiliary variable $\nu_{k,m}^{(\ell),[t+1]}$ is written as
\begin{align}
&\nu_{k,m}^{(\ell),[t+1]} = w_u\gamma_{k,u,m} -  w_u\sum\limits_{a \ne k}^{K_u}{d^{(\ell)}_{a,u,u,m}} - \eta^{(\ell),[t]}\xi_u  \ntb
&\quad + \sum\limits_{a \ne k }^{K_u} {\sum\limits_{n = 1}^{N_{a,u}} { \frac{w_u\left[\bOmega_{a,u,u}\right]_{n,m}} { \widetilde \gamma _{a,u,n} + \sigma^2 +C_{k,m,a,u}\left(\bLambda^{(\ell),[t]},\bLambda^{(\ell),[t+1]}\right) }} },
\end{align}
where
\begin{align}
&C_{k,m,a,u}\left(\bLambda^{(\ell),[t]},\bLambda^{(\ell),[t+1]}\right) = \sum_{q\neq u}^{U}\sum_{p=1}^{K_q} {\tr { \bLambda_{ p,q }^{(\ell),[t]} \widehat \bR_{a,u,q,n}}} \ntb
&\qquad + \sum\nolimits_{\left(a^{\prime}, m^{\prime}\right) \in \mathcal{S}_{k, m, a}} \lambda_{a^{\prime},m^{\prime},u}^{(\ell),[t+1]} \left[\bOmega_{a,u,u}\right]_{n,m^{\prime}},\\
&\mathcal{S}_{k, m, a}=\left\{\left(a^{\prime}, m^{\prime}\right) | a^{\prime} \neq a,\left(a^{\prime}, m^{\prime}\right) \neq(k, m), \right. \ntb
&\qquad \qquad \qquad \qquad \left. a^{\prime} \in\{1, \ldots, K_u\}, m^{\prime} \in\{1, \ldots, M\}\right\}.
\end{align}
This concludes the proof.

\end{appendices}


\end{document}